# Self-referenced spectral interferometry theory

T. Oksenhendler

[1] FASTLITE, Centre Scientifique d'Orsay, Bât. 503, Plateau du Moulon, BP45, 91401 Orsay, France,
email: thoksen@fastlite.com
tel : 33 1 69 35 87 10
fax : 33 1 69 35 88 44

Self-referenced spectral interferometry, a newly introduced ultrafast pulse characterization is described and reviewed theoretically. Validity range, temporal dynamic, resolution and precision are detailed in the scope of different experimental set-ups. Ultimate performances for ultrashort pulses in terms of minimal or maximal pulse durations, distortions, spectral bandwidth, temporal dynamic and range are estimated, simulated and explained in detail.

## 1    Introduction

As ultrashort pulses are shorter than any direct temporal characterization resolution, their temporal intensity and phase are measured indirectly. The complete determination of an ultrashort pulse requires to measure both its spectral amplitude and phase. The spectral phase is usually expanded into a Taylor's series whose absolute and first terms correspond to absolute phase and delay which are not relevant for pulse characterization.
As demonstrated by Wong and Walmsley[1], in the absence of any reference pulse, the complete temporal characterization of femtosecond pulses requires a nonlinear or non stationary filter.
The most widely used pulse measurement techniques, Frequency Resolved Optical Gating (FROG) [2] and Spectral Phase Interferometry for Direct Electric-field Reconstruction (SPIDER) [3], indeed rely on three-wave or four-wave mixing processes to generate a signal from which the spectral phase can be retrieved. Both these techniques are self-referenced and can be made single-shot by the use of non collinear harmonic generation. However, the algorithms used to retrieve the spectral complex amplitude are not straightforward. FROG belongs to the class of spectrographic measurements and relies on a blind iterative retrieval algorithm [4]. With the SPIDER technique, an analytic function of the spectral phase is directly measured but complete phase retrieval is obtained through an integration or concatenation step. Furthermore, it makes some assumptions on the pulse duration and spectral shape. Experimentally, as single-shot implementation of these techniques uses non collinear harmonic generation, their alignments are not straightforward.
The existence of a reference pulse, with a known spectral phase on a larger bandwidth than the pulse to measure, hugely simplifies the measurement setup and algorithm by using spectral interferometry [5,6]. This method is linear, analytic, sensitive and accurate. Unlike the SPIDER technique, neither shear nor harmonic generation is needed. However, to make this measurement self-referenced, the reference pulse has to be generated from the pulse itself.
In the Self-Referenced Spectral Interferometry (SRSI), the reference pulse is "self-created" from a temporal filter by a frequency-conserving nonlinear optical effect [7]. In the scope of this article, we consider third order frequency conservative effect: the Crossed-Polarized Wave generation (XPW) [8,9]. The creation mechanism and spectral characteristics of the self-created reference pulse are introduced in the next section. From this reference pulse, the method and the algorithm used to extract the input pulse spectral amplitude and phase are detailed step by step. In section 4, the validity range, convergence criteria and validity criteria are estimated analytically for purely chirped Gaussian pulses and simulated for other pulse shapes. Imperfections of experimental implementations are then reviewed in section 5 for optical setups examples, in section 6 for the third order non linear effect, and in section 7 for the spectrometer. Most of the distortions of the SRSI signal due to these imperfections can be corrected in the data processing described in section 8. In the final part, limitations of the method are estimated for some relevant examples: long (picosecond) and ultrashort (sub 10 fs) pulse measurements, temporal dynamic limits.

## 2 Reference pulse creation

In the spatial domain, an input beam can be spatially filtered to get rid of high frequency modulation in the Fourier domain. The output re collimated beam has a smoother and wider beam profile associated with a flatter spatial phase. The filter is directly obtained in a Fourier plane (focal plane) by inserting a hole. By analogy, a reference pulse in the spectral domain can be obtained by filtering the input pulse in the temporal domain. At the femtosecond time scale, the temporal filter can only be provided by a non linear effect. For reasons of compactness, simplicity, colinearity and achromaticity, the third order frequency conservative non linear effect used is the Cross-polarized Wave generation (XPW) [8, 9]. Other non-linear frequency conserving effects can also be used, such as the Self-diffraction [10].

XPW generated signal presents, to the first order, a cubic relationship with the time-dependent intensity of the input signal.

The analogy between spatial and temporal illustrates the filtering effect. In the spatial domain, an input beam corresponds to the spectral domain of an ultrashort pulse. This beam is focused by a lens. At the focal point, the spatial profile of the beam corresponds to the Fourier transform of the input beam (this plane is also named Fourier plane). In this analogy, this focal spot represents the temporal domain. The non linear filtering by XPW introduces a transmission that directly depends upon the intensity of the spot. If the focal spot is not too stretched, the filter transmission is equivalent to a pinhole. The filtering effect on the re collimated beam at the output is well known in the spatial domain to flatten the phase and enlarge the amplitude width. The same effect is expected in the spectral domain by the temporal filtering with XPW generation.

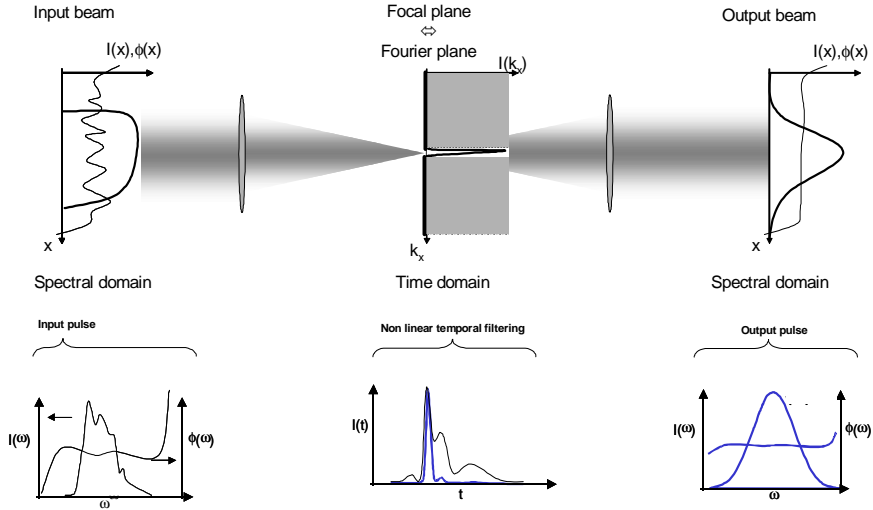

Fig.1: Temporal filtering effect illustration by spatio-temporal analogy.

For higher order spectral phase, the filtering is even more efficient. For chirp free pulses, duration shortening also broadens the spectrum. Thus, if an initial input pulse has no significant chirp, the XPW generated signal can be considered as a reference pulse with an approximately flat spectral phase.

This qualitative analysis can be analytically illustrated if we consider a specific pulse shape: a purely chirped Gaussian shape signal expressed as: $\tilde{E}^+(\omega) = e^{-\frac{(\omega-\omega_0)^2}{2\Delta\omega^2}} e^{-i\frac{1}{2}\phi^{(2)}(\omega_0)(\omega-\omega_0)^2}$

The mathematical description of the pulse and notation are those of the Femtosecond Laser Pulses of the Handbook of Lasers and Optics [11] and are reminded in Appendix A.

As demonstrated and illustrated by A.Jullien and al. [9], the cubic dependence of the third order non linear effect (XPW for example) reduces the input second order spectral phase (chirp) by a factor up to 9: it acts as a pinhole filtering in the time domain as shown on fig.1.

This non linear effect is considered as a first approximation as purely cubic:

$$E^+_{XPW}(t) \propto |E^+(t)|^2 E^+(t) \propto e^{-\frac{t^2}{\frac{2\Delta\tau_0^2}{3}\left(1+\left(\frac{\phi^{(2)}(\omega_0)}{\Delta\tau_0^2}\right)^2\right)}} e^{i6\Delta\tau_0^4\left(1+\left(\frac{\phi^{(2)}(\omega_0)}{\Delta\tau_0^2}\right)^2\right)t^2} \quad (1).$$

In the spectral domain, it is:

$$\tilde{E}^+_{XPW}(\omega) \propto e^{-\frac{(\omega-\omega_0)^2}{2\Delta\omega_{XPW}^2}} e^{-i\frac{1}{2}\phi^{(2)}_{XPW}(\omega_0)(\omega-\omega_0)^2} \quad (2).$$

Then for the theoretical XP Wave, the rms duration, rms spectral width and chirp are expressed as:

$$\Delta\tau_{XPW} = \frac{\Delta\tau_0}{Z(\Delta\tau_0, \phi^{(2)})} \Leftrightarrow \Delta\omega_{XPW} = Z(\Delta\omega_0, \phi^{(2)})\Delta\omega_0 \quad (3),$$

$$\phi^{(2)}_{XPW} = \frac{\phi^{(2)}}{3Z^2(\Delta\tau_0, \phi^{(2)})} = \frac{\phi^{(2)}}{Z_0^2 Z^2(\Delta\tau_0, \phi^{(2)})} \quad (4),$$

where $Z(\Delta\tau_0, \phi^{(2)}) = Z\left(x = \frac{\phi^{(2)}}{\Delta\tau_0^2}\right) = \sqrt{\frac{1}{3}\frac{9+(\alpha x)^2}{1+(\alpha x)^2}} = \sqrt{\frac{1}{Z_0^2}\frac{Z_0^4 + (\alpha x)^2}{1+(\alpha x)^2}}$ (5), expressed with the relative chirp factor x and where $\Delta\omega_0 \Delta\tau_0 = \alpha$.

Z factor corresponds to the pulse compression or spectral broadening by XPW generation. Its maximum is obtained for compressed pulse (x=0) $Z_{max} = Z_0 = \sqrt{3}$ and it decreases rapidly down to $Z = 1/\sqrt{3}$ for large chirps (|x|>>1).

For pulses that are compressed enough to have $Z > 1$, the spectral chirp in the XPW is less than a third of the initial chirp. In the same time, the spectrum of the XPW pulse is larger than the input pulse. These characteristics correspond to an ideal reference pulse for the Fourier transform spectral interferometry (FTSI) [5, 12].

### 3  Self-Referenced Spectral Interferometry Algorithm

The input pulse is compared to its non linearly filtered part by spectral interferometry (Fig.2). The principle is to compare both pulses in spectral amplitude and phase.

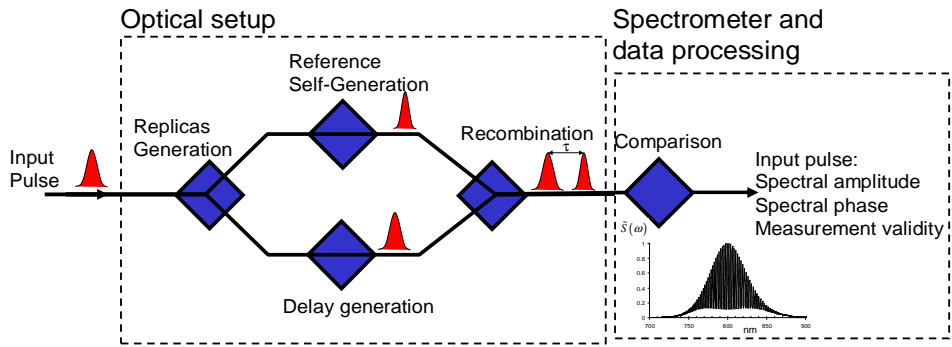

Fig.2 Self-referenced spectral interferometry principle scheme.

The measured interference spectrum can be expressed as

$$\tilde{S}(\omega) = \left|\tilde{E}^+_{XPW}(\omega) + \tilde{E}^+(\omega)e^{i\omega\tau}\right|^2$$
$$= \tilde{S}_0(\omega) + \tilde{f}(\omega)e^{i\omega\tau} + \tilde{f}^*(\omega)e^{-i\omega\tau} \quad (6),$$

where $\tilde{S}_0(\omega) = |\tilde{E}^+(\omega)|^2 + |\tilde{E}^+_{XPW}(\omega)|^2$ (7) is the sum of the spectra between the XPW and the input pulse to be measured, and $\tilde{f}(\omega) = \tilde{E}^+(\omega)\tilde{E}^{+*}_{XPW}(\omega)$ (8) is the interference part of the two pulses.

The two components of the interferogram, the DC term $\tilde{S}_0(\omega)$ and the AC term $\tilde{f}(\omega)$, result from classical FTSI processing as shown on fig.3.

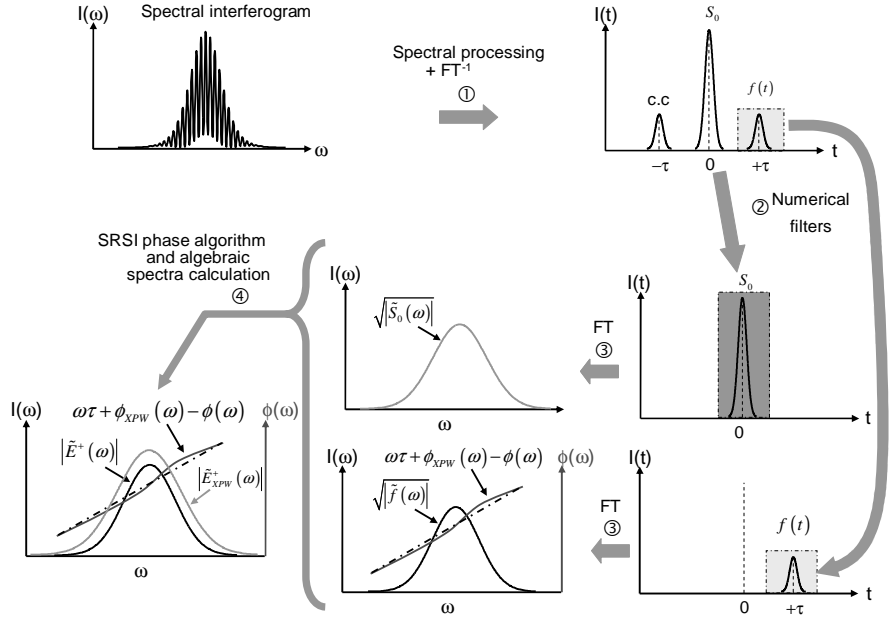

Fig.3 FTSI processing to extract $\tilde{S}_0(\omega)$ and $\tilde{f}(\omega)$ from the measured interferogram (steps 1 to 3) and SRSI phase algorithm and algebraic spectra calculation (step 4).

Under the condition that the XPW spectral components are more intense than input pulse ones at each wavelength, the spectra of the two pulses are analytically calculated by the following expressions:

$$|\tilde{E}^+_{XPW}(\omega)| = \frac{1}{2}\left(\sqrt{\tilde{S}_0(\omega)+2|\tilde{f}(\omega)|} + \sqrt{\tilde{S}_0(\omega)-2|\tilde{f}(\omega)|}\right)$$
$$|\tilde{E}^+(\omega)| = \frac{1}{2}\left(\sqrt{\tilde{S}_0(\omega)+2|\tilde{f}(\omega)|} - \sqrt{\tilde{S}_0(\omega)-2|\tilde{f}(\omega)|}\right)$$ (9).

The spectral phase is first estimated by the argument of $\tilde{f}(\omega)$, considering initially that the XPW pulse spectral phase is null or at least negligible compared to the input pulse spectral phase:

$$\phi(\omega) = \phi_{XPW}(\omega) - \arg(\tilde{f}(\omega)) \approx -\arg(\tilde{f}(\omega))$$ (10).

From this rough estimation, the input pulse temporal profile and the XPW pulse can be simulated giving a first estimation of the XPW pulse spectral phase. This phase is then re-introduced in the expression (10) and these steps repeated until the phase modification is negligible. This iterative algorithm is described on fig. 4.

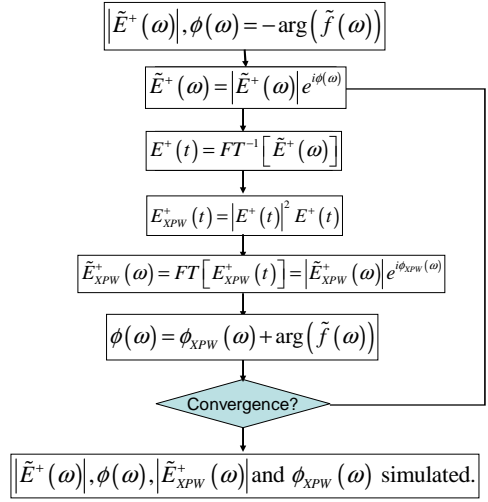

Fig.4: Iterative algorithm for precise spectral phase determination.

From the complete algorithm, the XPW spectrum is analytically calculated directly from the measured signal and also simulated from the input pulse determination. The input spectrum is also analytically calculated and its spectral phase estimated through the iterative algorithm as long as it converges.

## 4  Algorithm convergence, validity range and criteria

The algorithm convergence condition corresponds to the validity range of the measurement. As assumed from the analysis of the XPW filtering effect, the validity of the measurement is directly linked to the quality of the temporal filter.

Knowing the input pulse spectrum, one should want to know a priori the validity range of the measurement, if the pulse can be measured.

Experimentally, once the measurement is done, a posteriori validity criteria are important to assess the precision quality of the measurement. In this case, validity criteria issued directly from the SRSI algorithm results are required.

### *4.1  Analytical analysis for Purely chirped Gaussian pulses*

#### *4.1.1  Validity range*

As mention in section 2, for a Gaussian pulse, the XPW filtering effect can analytically be expressed by the broadening factor:

$$Z(\Delta\tau_0, \phi^{(2)}) = Z\left(x = \frac{\phi^{(2)}}{\Delta\tau_0^2}\right) = \frac{\Delta\tau_0}{\Delta\tau_{XPW}} = \frac{\Delta\omega_{XPW}}{\Delta\omega_0} = \sqrt{\frac{1}{3}\frac{9+(\alpha x)^2}{1+(\alpha x)^2}}$$ (11), where x is the relative rms chirp factor

of the input pulse and $\alpha = \alpha_0 = \Delta\omega_0 \Delta\tau_0 = \frac{1}{2}$ the time-bandwidth product of the Fourier limited pulse for a

Gaussian pulse. Its maximum value is: $Z_0 = \max(Z(x)) = Z(0) = \frac{\Delta\omega_{0XPW}}{\Delta\omega_0} = \sqrt{3}$.

The complete SRSI algorithm can then be treated as a sequence of (x). The first estimation of the spectral phase, directly resulting from the measurement and FTSI data processing, is the first term of the sequence

$$x_m = x_0 = \frac{\arg(\tilde{f}(\omega))}{\Delta\tau_0^2}$$ (12).

For a pure Gaussian chirped pulse, the XPW pulse chirp can be expressed as $x_{XPW} = x\left(\dfrac{1+(\alpha x)^2}{9+(\alpha x)^2}\right)$ (13), the second term of the sequence is: $x_1 = x_0 + x_0\left(\dfrac{1+(\alpha x_0)^2}{9+(\alpha x_0)^2}\right)$ (14).

By iteration, the i+1th term is expressed as: $x_{i+1} = x_0 + x_i\left(\dfrac{1+(\alpha x_i)^2}{9+(\alpha x_i)^2}\right)$ (15).

The SRSI technique is valid if and only if this sequence converges to the input pulse relative chirp: $\lim(x) = x_\infty = x$.

As the sequence is defined by recurrence, its convergence condition and limit depend upon the initial term. It is convergent if and only if: $|x_0| \leq \dfrac{4}{3\alpha}$, and the limit is: $x_\infty = \dfrac{4 - \sqrt{16 - 9(\alpha x_0)^2}}{\alpha^2 x_0}$ (16).

From the measurement interest it has to be expressed from the input relative chirp x.

The initial term corresponds to the output from the SRSI measurement: $x_m = x_0 = x - x\left(\dfrac{1+(\alpha x)^2}{9+(\alpha x)^2}\right) = \dfrac{8x}{9+(\alpha x)^2}$

(17). This function has extrema for $x = \pm\dfrac{3}{\alpha}$, while its values are in the range $\left[-\dfrac{4}{3\alpha};\dfrac{4}{3\alpha}\right]$ (fig.5 $x_0$ versus x).

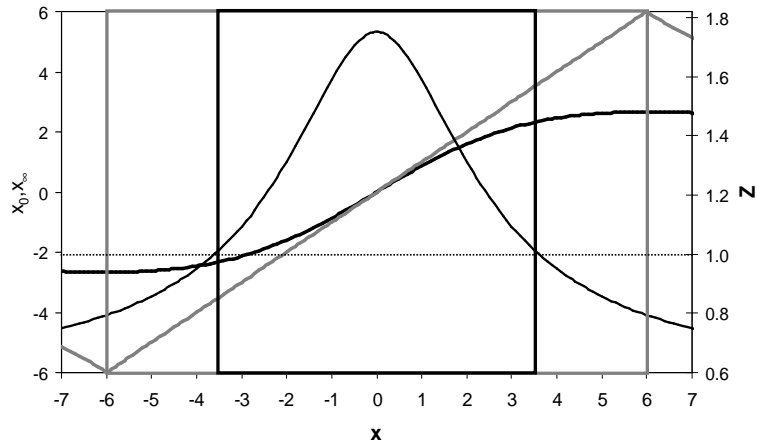

Fig.5 : $x_\infty$ (thick grey curve), $x_0$ (thick black curve) and Z (thin black curve) versus x

Thus in any experimental case, the sequence of the algorithm is converging.

But the validity of the measurement also requires that the limit $x_\infty$ of the sequence is x. As shown on fig.5 by the curve $x_\infty$ versus x, $x_\infty = x$ only if $x \in \left[-\dfrac{3}{\alpha};\dfrac{3}{\alpha}\right]$.

This interval represented on fig.5 by a grey rectangle is the theoretical validity range of the measurement.

Experimentally, the algorithm is efficient if its rate of convergence is high enough and if the initial term variation versus the input pulse relative chirp ($\frac{dx_0}{dx}$) is sufficient to overcome fluctuations or other defaults that will be described hereafter.

The rate of convergence decreases as $x_0 \to \frac{4}{3\alpha}$, when $\frac{dx_0}{dx} \to 0$.

A rule of thumb is defined as $\frac{dx_0}{dx} \geq \frac{1}{3} \Leftrightarrow |x| \leq \frac{\sqrt{3}}{\alpha}$. Within this interval the rate of convergence is very fast: 15 iterations lead to a precision better than 0.1% on x.

Furthermore this interval can be easily check experimentally by the broadening factor Z as it corresponds to Z>1. It is represented on figure 5 by a black rectangle, named hereafter "conservative validity range".

Inside this range the measurement is valid and efficient. It is estimated directly from the input pulse spectrum by $\alpha$ and $Z_0$.

*4.1.2 Validity criteria*

The validity range determines the ability of the method to effectively measure an a priori distorted pulse. It differs from a validity criteria that should assess the measurement quality a posteriori.

This difference can be highlighted by a simple example. Lets consider an input pulse with a large chirp (x=12). The initial term is then $x_0 \approx 2.13$ and the algorithm converges to $x_\infty \approx 3$. The measurement is totally inaccurate ($x_\infty \neq x$).

In this case, the input pulse chirp is out of the validity range defined previously. But the result of the measurement is completely inside this range. From the user, it is not possible to determine uniquely with the phase result, the quality of the measurement.

Furthermore, whatever is the input pulse chirp, the initial term is always inside the theoretical validity range: $\left[-\frac{4}{3\alpha}; \frac{4}{3\alpha}\right] \subset \left[-\frac{3}{\alpha}; \frac{3}{\alpha}\right]$. Thus the algorithm always converges onto a result $x_\infty$.

Additional criteria are required to assess the validity of measurement i.e. $x = x_\infty$.

To represent the algorithm convergence, it is interesting to compare the input relative chirp factor $x$ with the limit of the algorithm sequence $x_\infty$. By adding the visualization of the broadening factor $Z$, the validity range is also shown. The curves of interest are then $Z$ versus $x$ compared to $Z$ versus $x_\infty$ (figure 6).

Inside the validity range, one expects that $x = x_\infty$ and thus the curves $Z$ versus $x$ and $Z$ versus $x_\infty$ are superposed.

Outside of the grey area, i.e. for input relative chirp factor $x$ greater than $\frac{3}{\alpha}$, the input pulse absolute relative chirp factor first estimation $x_0$ is still in the range $\left[-\frac{4}{3\alpha}; \frac{4}{3\alpha}\right] \subset \left[-\frac{3}{\alpha}; \frac{3}{\alpha}\right]$ as illustrated by the previous example. This algorithm cannot be used. The curves $Z$ versus $x$ and $Z$ versus $x_\infty$ are not superposed anymore (when $x \to \infty$, $x_\infty \to 0$).

As the SRSI method outputs also the input and XPW pulses spectra. The effective broadening factor can be determined: $Z(x) = \frac{\Delta\omega_{XPW}}{\Delta\omega_0}$. As mention before, the conservative range of validity matches exactly the broadening condition $Z > 1$. This criterion can be determined a posteriori from experimental results. Is this criterion sufficient? In the previously given example with x=12, the output of the SRSI gives two spectra. One should pay attention that, in this case, the initial pulse can be mistaken as the XPW pulse (larger spectral width). The compression factor estimated is then the inverse. As shown on fig.6, the situation is symmetric for Z and 1/Z with axis of symmetry Z=1. The validity criteria of the broadening are apparently fulfilled and the measurement seems accurate despite its wrong results.

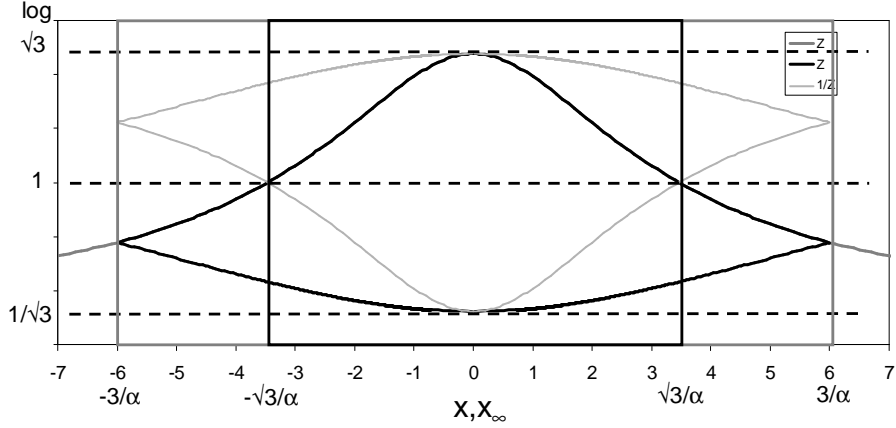

Fig.6: symmetric behavior of initial pulse and XPW pulse: Z versus $x_\infty$ (thick black curve), Z versus x (thick black curve with cross markers) and 1/Z versus $x_\infty$ (thin black curve).

To avoid this measurement error, a comparison with the initial pulse spectrum measured separately is required. No confusion on which spectrum is the initial one can then be made. No confusion between Z and 1/Z can be done and the symmetry is broken.

Thus the validity criteria are the broadening factor $Z > 1$ and the input pulse spectrum compared with an initial separate measurement. An additional qualitative factor can be given by the comparison of the XPW pulse spectra recovered both from the measurement analytically and by simulation through the algorithm.

To sum up the purely Chirped Gaussian pulse SRSI measurement study, as a rule of thumb, the measurement is valid in a conservative validity range as long as $Z > 1$ i.e. $|x| < \frac{Z_0}{\alpha} = 2\sqrt{3} \approx 3.46$. The validity of the measurement can be assessed by comparing the separately measured input spectrum with the recovered pulse spectrum.

### 4.2 *Beyond the scope of the Gaussian pulse*

The simple analytical treatment of SRSI algorithm can only be applied to purely chirped Gaussian pulse. Unfortunately, real ultrafast laser spectra usually differ from a Gaussian shape and higher order spectral phase terms are also of interest in the pulse characterization. In this part, the validity range and validity criteria given for purely chirped Gaussian pulses will be extended for general pulse shapes and higher order spectral phases.

As previously mentioned, the validity range has to be determined from the input pulse spectrum only. It is indeed useful to know a priori if this method can be applied to an experiment or not.

On the other hand, the results of the measurement have to be checked for their validity and precision. As detailed in section 3, the results of the SRSI method are: the input pulse spectrum $\left|\tilde{E}^+(\omega)\right|^2$, the XPW pulse spectrum $\left|\tilde{E}^+_{XPW}(\omega)\right|^2$, the spectral phase of the input pulse $\phi(\omega)$, the spectral phase of the XPW pulse $\phi_{XPW}(\omega)$, the simulated XPW pulse spectrum $\left|\tilde{E}^+_{XPW}(\omega)\right|^2_{simulated}$.

The purely chirped Gaussian pulse is specific because it is also a purely chirped Gaussian pulse in the time domain and even through the non linear temporal filter. The effect of the temporal filter is easily estimated in this case and depends mainly upon the broadening factor. This factor differs with other shape or higher orders. On the other hand, the time-bandwidth product, also involved in the validity range, is minimal for Gaussian shape: $\alpha_0 = \Delta\tau_0 \Delta\omega = 0.5$. The non linear temporal filter effect cannot thus decrease this time-bandwidth product: $\alpha_{XPW} = \Delta\tau_{XPW}\Delta\omega_{XPW} = \alpha_0 = \Delta\tau_0\Delta\omega = 0.5$.

As the quality of the measurement is directly due to the temporal filter, the non linear filter efficiency should be estimated either a priori from the input pulse spectrum to determine the validity range for different spectral phase orders or a posteriori from SRSI results to assess the validity of the measurement done.

*4.2.1 Validity range*

4.2.1.1 Non Gaussian chirped pulses

Chirped Pulse Amplification lasers more generally use a Treacy grating compressor. The finite size of the gratings sharply limits the bandwidth. The spectral shape is closer to super Gaussian or rectangular than Gaussian. In order to fit more realistic femtosecond pulse spectra, the two spectra considered hereafter are:

- Spectrum 1: "super Gaussian" of order 3 : $\left|\tilde{E}^+(\omega)\right| = e^{-\frac{1}{2}\left(\frac{\omega-\omega_0}{\Delta\omega}\right)^6}$, with $\omega_0 = {2\pi c}/{800 nm}$ the central pulsation and $\Delta\omega = 2\pi c \left(80/800^2\right) nm^{-1}$ the width,

- Spectrum 2: asymmetric "super Gaussian" of order 3 with a Gaussian hole: $\left|\tilde{E}^+(\omega)\right| = \left(1 - H_{depth} e^{-\frac{1}{2}\left(\frac{\omega-\omega_{Hole}}{\Delta\omega_H}\right)^2}\right) e^{-\frac{1}{2}\left(\frac{\omega-\omega_0}{\Delta\omega}\right)^6}$, where $H_{depth} = 0.9$ is the hole depth, $\Delta\omega_H = 2\pi c\left(25/800^2\right) nm^{-1}$ is the hole width, $\omega_{Hole} = {2\pi c}/{795 nm}$ is the central position of the hole, $\Delta\omega = 2\pi c\left(80/800^2\right) nm^{-1}$ the width and $\omega_0 = {2\pi c}/{800 nm}$ the central pulsation of the super Gaussian part.

Our goal in this part is to extrapolate the validity ranges determined for purely chirped Gaussian pulses :

- the maximal validity range : $\frac{dx_0}{dx} \geq 0 \Leftrightarrow |x| \leq \frac{Z_0^2}{\alpha}$,

- the rule of thumb corresponding to the conservative validity range : $\frac{dx_0}{dx} \geq \frac{1}{3} \Leftrightarrow |x| \leq \frac{\sqrt{3}}{\alpha} = \frac{Z_0}{\alpha}$.

From the input spectrum, one can simulate numerically the SRSI first estimation to a relative chirp x. By scanning the relative chirp, one should calculate the derivative $\frac{dx_0}{dx}$ values and thus the validity ranges. This operation is fastidious and one should estimate approximately the validity range by a rule of thumb.

From the input spectrum, one can calculate digitally the Fourier limited pulse temporal intensity. The rms duration $\Delta\tau_0$ defined as the statistical width (standard deviation) is then calculated on this temporal intensity. The minimum time-bandwidth product defined as $\alpha = \alpha_0 = \Delta\tau_0\Delta\omega_0$ is obtained for this pulse spectrum.

From the Fourier limited pulse temporal intensity, one can simulate its XPW pulse spectrum. The rms spectral width of this XPW $\Delta\omega_{0XPW}$ and the time-bandwidth product of the XPW pulse are then calculated as this pulse is by hypothesis Fourier transform: $\alpha_{XPW} = \Delta\tau_{0XPW}\Delta\omega_{0XPW}$ (18).

By opposition to the Gaussian pulse case, the time-bandwidth products are not equal: $\alpha_{XPW} \neq \alpha$ and not minimal: $\alpha > \alpha_{XPW} > 0.5$.

The maximum compression factor defined as the ratio of the two spectral widths can also be digitally estimated: $Z_0 = Z_{0\omega} = \Delta\omega_{0XPW}/\Delta\omega_0$. This compression factor characterizes the pulse shape in comparison with a Gaussian pulse.

The deviation of the time-bandwidth product of the XPW pulse to the minimal value means the deviation of the filtered pulse to its optimal shape, the Gaussian shape. The temporal filtering can be seen from the statistics point of view as a combination of n photons (three in our case). The central limit theorem states that as n gets larger, the distribution (intensity in here) approximates normal (Gaussian shape) with the same mean and a variance divided by n. In our case, it means that the XPW pulse approximates a Gaussian pulse with the same position and a width divided by $\sqrt{3}$. But as 3 is not large enough to consider that the approximation is fully valid, this estimation needs to be adapted. One should note that the factor $\alpha_{XPW}$ should be closer to 0.5 than $\alpha$ meaning that the XPW pulse is closer to a Gaussian and that the filter is efficient.

As the non linear filter outputs a pulse relatively close to a Gaussian shape, the previous demonstration with Gaussian shape pulses can be re used by defining equivalent input Gaussian pulses (described by subscript G).

For the non linear pulse one can define an equivalent Gaussian pulse with spectral width: $\Delta\omega_{0GX} = \Delta\omega_{0XPW}$.

For the input pulse, one can define a Gaussian pulse whose non linear result is this equivalent non linear Gaussian pulse. The width of this pulse can be expressed as: $\Delta\omega_{0G} = \frac{\Delta\omega_{0GX}}{\sqrt{3}} = \frac{\Delta\omega_{0XPW}}{\sqrt{3}} = \frac{Z_0\Delta\omega_0}{\sqrt{3}}$ (19), and its chirps by: $\phi_{0G}^{(2)}\Delta\omega_{0G}^2 = \phi_0^{(2)}\Delta\omega_0^2$. It implies: $\phi_{XPW}^{(2)} = \frac{\phi_{0G}^{(2)}}{3Z^2(x)} = \frac{\phi^{(2)}}{Z_0^2 Z^2(x)}$ (20).

It follows from the previous expression for the Gaussian pulse:

$$Z(x) = \sqrt{\frac{1}{Z_0^2}\frac{Z_0^4 + (\alpha x)^2}{1 + (\alpha x)^2}}, \quad x_{XPW} = x\frac{1 + (\alpha x)^2}{Z_0^4 + (\alpha x)^2}$$

and the initial signal from the SRSI:

$$x_m = x_0 = x\left(\frac{Z_0^4 - 1}{Z_0^4 + (\alpha x)^2}\right).$$

The expression of the validity range and the rule of thumb of the conservative validity range can then be expressed as:

- $-\frac{dx_0}{dx} \geq 0 \Leftrightarrow |x| \leq \frac{Z_0^2}{\alpha}$ (21)

- $-\frac{dx_0}{dx} \geq \frac{1}{3} \Leftrightarrow |x| \leq \frac{\sqrt{\sqrt{33Z_0^8 - 42Z_0^4 + 9} - 5Z_0^4 + 3}}{\sqrt{2}\alpha} = \frac{B}{\alpha}$ (22).

One should note that the conservative range can be null. Indeed, the maximum value for $\frac{dx_0}{dx}$ is obtained for x=0:

$\left.\frac{dx_0}{dx}\right)_{x=0} = \frac{Z_0^4 - 1}{Z_0^4}$ (23). This value is positive but can be lower than 1/3 if $Z_0 < 1.1$. In this case, the conservative validity range doesn't exist.

Spectrum 1 is a super Gaussian in frequency domain. Its minimum time-bandwidth product is $\alpha \approx 0.71$. Its maximum compression factor is: $Z_0 = 1.43$.

For spectrum 2 these parameters are: $\alpha \approx 1.52, Z_0 = 1.37$.

No analytical analysis can be done on such pulses. The computer simulation implemented emulates the spectrum and uses the SRSI algorithm to recover the spectral phase. By sweeping the chirp, the curve of $Z$ is calculated versus the relative phase factor x introduced and $x_0$ measured by the SRSI algorithm.

The number of loops of the iterative algorithm is limited to 15.

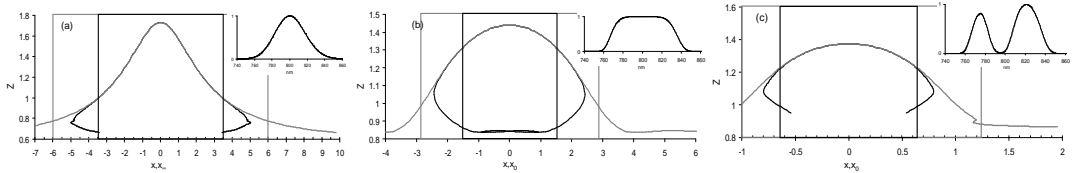

Fig.7: (a) Gaussian shape with second order spectral phase ($X_{max} = 6$, $X_{lim} = 3.46$), (b) spectrum 1 with second order spectral phase ($X_{max} \approx 2.8$, $X_{lim} \approx 1.5$) and (c) spectrum 2 with second order spectral phase ($X_{max} \approx 1.2$, $X_{lim} \approx 0.63$).

4.2.1.2 Higher order spectral phase pulses

For higher order spectral phase terms, considering the order n of the spectral phase Taylor development, a more general relative spectral phase factor is introduced:

$x^{(n)} = \frac{\phi^{(n)}}{\Delta\tau_0^n}$ (24).

The rms duration is expressed as:

$\Delta\tau_g^2\left(\phi^{(n)}\right) = \int \left(\frac{\partial\phi}{\partial\omega}\right)^2 |\tilde{E}^+(\omega)|^2 \frac{d\omega}{2\pi} \Big/ \int |\tilde{E}^+(\omega)|^2 \frac{d\omega}{2\pi} = \left(\phi^{(n)}/(n-1)!\right)^2 \left\langle(\omega-\omega_0)^{2(n-1)}\right\rangle$ (25).

We define: $\left\langle(\omega-\omega_0)^{2n}\right\rangle = \beta_{2n}\sigma_\omega^{2n} = \beta_{2n}\Delta\omega^{2n}$ (26).

Thus, $\Delta\tau_g^2\left(\phi^{(n)}\right) = \beta_{2(n-1)}\left(\phi^{(n)}/(n-1)!\right)^2 \left(\Delta\omega_{rms}^{2(n-1)}\right) = \Delta\tau_0^2 \left(\left(\sqrt{\beta_{2(n-1)}}\alpha'^{(n-1)}/(n-1)!\right)\frac{\phi^{(n)}}{\Delta\tau_0^n}\right)^2$ (27).

In statistics, $\beta_n$ is equal to 1 for n=2, named "skewness" for n=3 and "kurtosis" for n=4. This factor is dependant upon the pulse shape.

Thus by introducing the relative spectral phase factor $x^{(n)}$ :

$\Delta\tau\left(x^{(n)}\right) = \sqrt{\Delta\tau_0^2 + \Delta\tau_g^2} = \Delta\tau_0\sqrt{1+\left(\alpha'^{(n-1)}\sqrt{\beta_{2(n-1)}}x^{(n)}/(n-1)!\right)^2} = \Delta\tau_0\sqrt{1+\left(\alpha'_n x^{(n)}\right)^2}$ (28).

The previously obtained validity ranges $|x| \leq X_{\lim} = \dfrac{\sqrt{\sqrt{33Z_0^8 - 42Z_0^4 + 9} - 5Z_0^4 + 3}}{\sqrt{2}\alpha} = \dfrac{B}{\alpha}$ and $|x| \leq X_{\max} = \dfrac{Z_0^2}{\alpha}$ needed to be adapted.

The term $\alpha$ is replaced by $\alpha_n = \dfrac{(\alpha)^{n-1}\sqrt{\beta_{2(n-1)}}}{(n-1)!}$ (29).

Thus for high orders spectral phase terms, the validity ranges are:

- $\left|x^{(n)}\right| \leq X_{\lim} = B/\alpha_n = \left((n-1)!B\right)/\left((\alpha)^{n-1}\sqrt{\beta_{2(n-1)}}\right)$ (30) is the conservative range represented by black rectangle on the figures of Z versus x and $x_0$,

- $\left|x^{(n)}\right| \leq X_{\max} = Z_1^2/\alpha_n = \left((n-1)!Z_1^2\right)/\left((\alpha)^{n-1}\sqrt{\beta_{2(n-1)}}\right)$ (31) is the maximum range of convergence of the algorithm illustrated by a grey rectangle on the figures.

Using the Gaussian analogy, one can approximate compared to the statistics (exact for n=1 and 2 but under estimated for n>2) $\beta_{2n}$ as: $\beta_{2n} = \left(\dfrac{Z_0}{\sqrt{3}}\right)^{2n}(2n-1)!!$, where $(2n-1)!! = \prod_{j=1}^{n}(2j-1)$. For first orders, $\beta_2 = \left(\dfrac{Z_0}{\sqrt{3}}\right)^2$, $\beta_4 = 3\left(\dfrac{Z_0}{\sqrt{3}}\right)^4$, $\beta_6 = 15\left(\dfrac{Z_0}{\sqrt{3}}\right)^6$. The validity ranges are for order 3 and 4: $\left|x^{(2)}\right| \leq 2\sqrt{3} \leq 6$, $\left|x^{(3)}\right| \leq 8 \leq 8\sqrt{3}$ and $\left|x^{(4)}\right| \leq 21 \leq 37$.

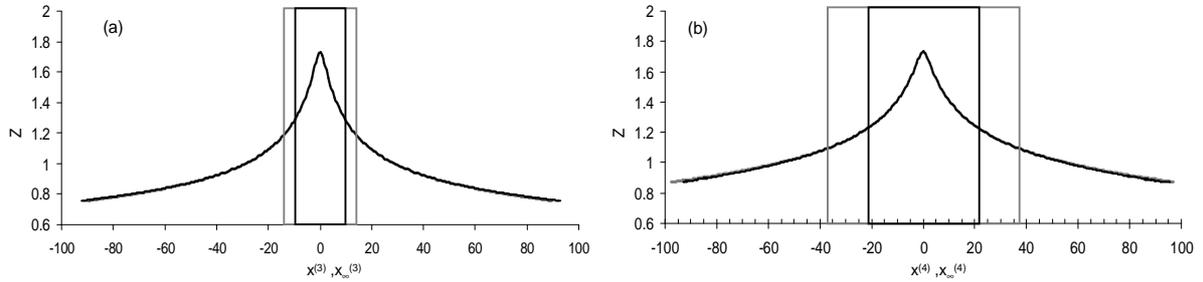

Fig. 8 : Validity ranges for Gaussian pulse shape, Z versus x and $x_\infty$ with (a) third order spectral phase (b) fourth order spectral phase.

As expected from the XPW non linear filtering part, higher orders are well filtered because the induced pulse distortions are rather low in temporal amplitude. The validity ranges are very conservative.

For spectra 1 and 2 with orders 3 and 4 the validity ranges are shown on figures 9 and 10.

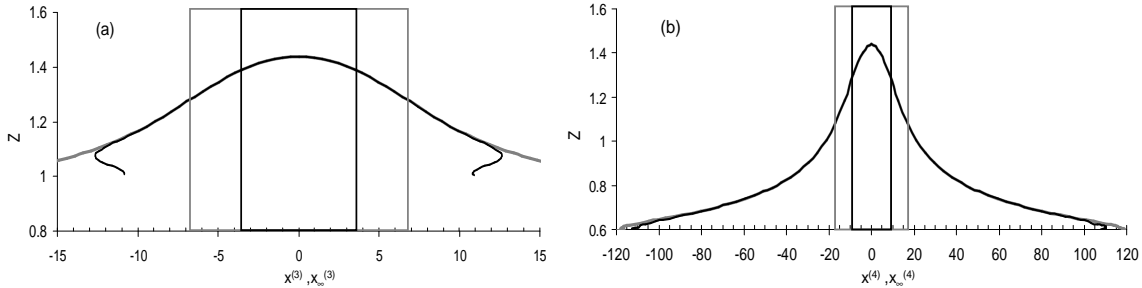

Fig. 9: Validity ranges for spectrum 1, Z versus x and $x_\infty$ with (a) third $\left|x^{(3)}\right| \leq 3.6 \leq 6.9$ and (b) fourth order $\left|x^{(4)}\right| \leq 8.6 \leq 17$.

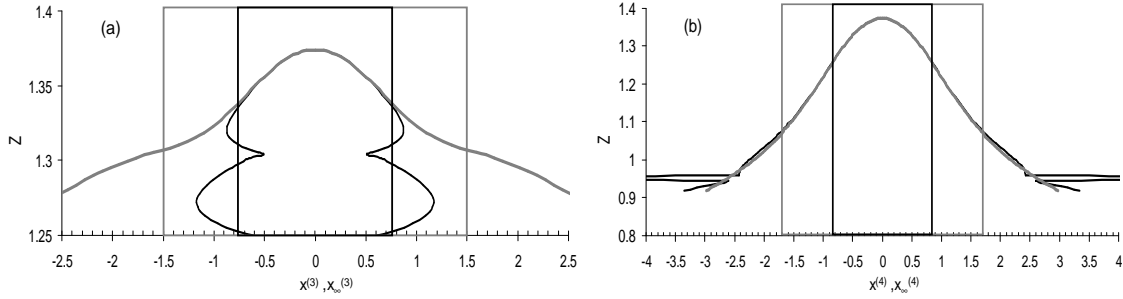

Fig. 10: Validity ranges for spectrum 2, Z versus x and $x_\infty$ with (a) third $\left|x^{(3)}\right| \leq 0.76 \leq 1.5$ and (b) fourth order $\left|x^{(4)}\right| \leq 0.85 \leq 1.7$.

For more general spectral phase profile, the Taylor's development can be used:

$$\phi(\omega) = \sum_{j=0}^{\infty} \frac{\phi^{(j)}(\omega_0)}{j!} (\omega - \omega_0)^j \text{ with } \phi^{(j)}(\omega_0) = \left.\frac{\partial^j \phi(\omega)}{\partial \omega^j}\right|_{\omega_0} \quad (5).$$

The first two terms have no influence on the temporal profile of the pulse. The higher order terms will modify the group delay of the pulse as:

$$\Delta \tau_g^2(\phi) = \int \left(\frac{\partial \phi}{\partial \omega}\right)^2 \left|\tilde{E}^+(\omega)\right|^2 \frac{d\omega}{2\pi} \bigg/ \int \left|\tilde{E}^+(\omega)\right|^2 \frac{d\omega}{2\pi} = \sum_{n=2}^{+\infty} \left(\phi^{(n)}/(n-1)!\right)^2 \left\langle (\omega-\omega_0)^{2(n-1)} \right\rangle \quad (32).$$

Using the relation (26) we finally obtain a relation involving only the spectral width of the pulse:

$$\Delta \tau_g(\phi) = \sqrt{\sum_{n=2}^{+\infty} \left(\phi^{(n)}/(n-1)!\right)^2 \beta_{2(n-1)} \Delta\omega^{2(n-1)}} = \Delta\tau_0 \sqrt{\sum_{n=2}^{+\infty} \left(\phi^{(n)}/(n-1)!\right)^2 \beta_{2(n-1)} \frac{\alpha^{2(n-1)}}{\Delta\tau_0^{2n}}} = \Delta\tau_0 x \quad (33).$$

The validity range are then defined as:

$$X_{\max} \equiv \phi_{\max} \Leftrightarrow \sum_{n=2}^{+\infty} \left(\phi^{(n)}/(n-1)!\right)^2 \beta_{2(n-1)} \frac{\alpha^{2(n-1)}}{\Delta\tau_0^{2n}} = Z_1^2 \sim \quad (34)$$

and $X_{\lim} \equiv \phi_{\lim} \Leftrightarrow \sum_{n=2}^{+\infty} \left(\phi^{(n)}/(n-1)!\right)^2 \beta_{2(n-1)} \frac{\alpha^{2(n-1)}}{\Delta\tau_0^{2n}} = B \quad (35).$

One should note that introducing the "relative relative" chirp factor $X = \frac{x}{X_{\lim}} = \frac{x}{B}$ normalizes the conservative validity range to $X \in [-1;1]$ for any shape and any spectral phase terms combination.

As this factor is not directly linked to usual parameters such as spectral phase order and duration of the pulse, in the following of this article, the relative spectral phase factor will be used.

As illustrated by these examples, the conservative validity range is robust even for rather distorted spectra. It is very conservative for Gaussian like pulse shapes. But it cannot be robust for any kind of pulses. In case of very complex structures, and if it is required, it is possible to digitally simulate the convergence of the algorithm as done here.

These results are computer simulations of the SRSI method. Some artifacts may seem to result from the simulation but are indeed limitations of the measurement.

One may note that even if the pulse to measure is out of the validity range, the spectral phase out of the measurement is a part of this input phase. This feature is very important for the pulses optimization feedback loop. It ensures that on a very large range, a feedback loop with a pulse shaper will converge to the perfectly compressed pulse. Out of range, the convergence rate of the feedback loop is then smaller than for in range pulses.

*4.2.1 Validity criteria*

From the measured spectrum, the XPW pulse of the Fourier limit is calculated and its widths estimated. The maximal spectral broadening factor $Z$ is then calculated as the ratio of the rms spectral width of this XPW Fourier limit pulse and the measured spectrum rms spectral width. The value of $Z$ at the validity range limit is then expressed as:

$$Z_{\text{limit}} = \sqrt{\frac{1}{Z_0^2} \frac{Z_0^4 + (B)^2}{1 + (B)^2}} \quad (35), \text{ where } B = \frac{\sqrt{\sqrt{33Z_0^8 - 42Z_0^4 + 9} - 5Z_0^4 + 3}}{\sqrt{2}}.$$

The validity criteria are then:

1. Comparison of the input pulse spectral intensity with a previously measured one (to avoid wrong measurements where the XPW pulse and input pulse are exchange for the algorithm c.f. fig.6),

2. Checking that the value of $Z$ estimated with the two measured spectra is higher than $Z_{\text{limit}}$ : $Z > Z_{\text{limit}}$,

3. Checking that the value of $Z$ estimated with the XPW simulated and measured input pulse spectra is higher than $Z_{\text{limit}}$ : $Z_{simulated} > Z_{\text{limit}}$,

4. Qualitative checking on the similarity between the XPW measured and simulated spectra.

As an example these criteria are estimated for pulses with spectrum 2 in the range of validity and out of the range of validity. The XPW measured and simulated spectra are shown on the figure 11 below:

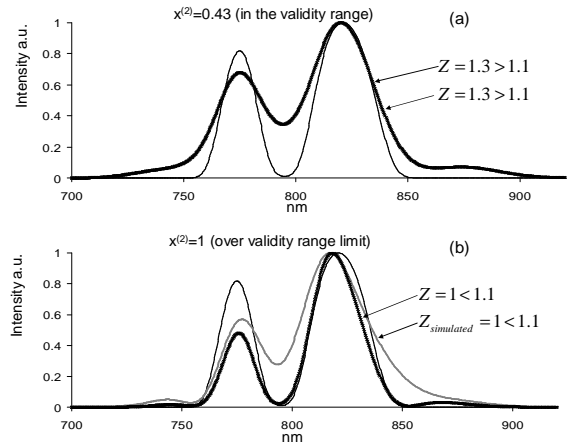

Fig.11: spectra of the input pulse and XPW measured and simulated with validity criteria for precise measurement and out of range measurement for spectrum 2 where $Z_{\text{limit}} = 1.1$ with (a) $Z = 1.3$ and (b) $Z = 1$.

As expected, if the validity criteria is not fulfilled, the measurement is inexact as illustrated by the difference between the measured and simulated XPW spectra in this simulation.

*4.3 Algorithm convergence and theoretical measurement precision*

The convergence of the algorithm has been assessed by direct comparison of the spectral phase Taylor's polynomial coefficients. This method has strong limitation for more complex spectral phases and cannot be generalized. As proposed and discussed by Dorrer and Walmsley [13], the pulse measurement precision should be characterized by a root-mean-square variation defined as the "rms precision error":

$$\varepsilon = \left\| \tilde{E}_1^+ - \tilde{E}_2^+ \right\| = \left[ \int_{-\infty}^{\infty} \left| \tilde{E}_1^+ - \tilde{E}_2^+ \right|^2 \frac{d\omega}{2\pi} \right]^{\frac{1}{2}} \quad (36) \quad \text{where the pulse energies are normalized :}$$

$$\left\| \tilde{E}_1^+ \right\| = \int_{-\infty}^{\infty} \left| \tilde{E}_1^+ \right|^2 \frac{d\omega}{2\pi} = 1 = \left\| \tilde{E}_2^+ \right\|.$$

By using $\tilde{E}_1^+$ the initial pulse spectrum and $\tilde{E}_2^+$ the measured pulse spectrum by SRSI, the precision of the measurement can be extended over pure polynomial spectral phases. This rms precision error is 0 for perfect measurements, 2 for completely wrong measurements. For the authors a good measurement is expressed as $\varepsilon < 0.1$ and an excellent one as $\varepsilon < 0.02$.

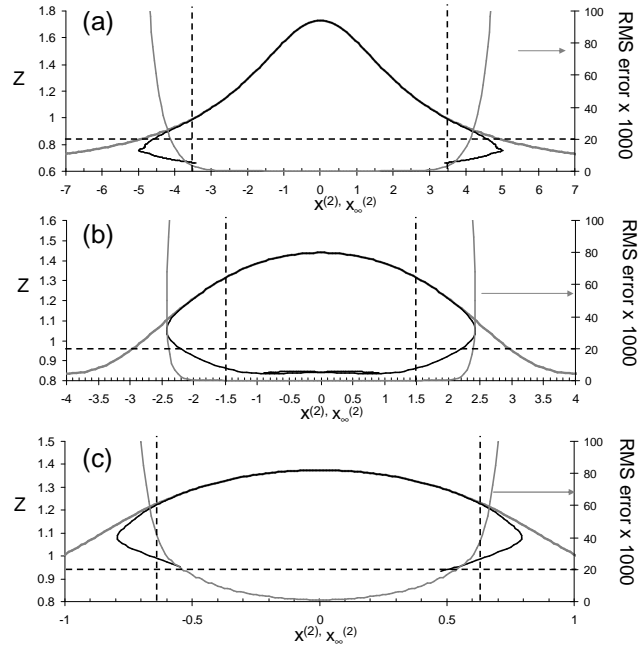

Fig.12 . Z versus x and $x^\infty$ , rms precision error versus x for chirped Gaussian and super Gaussian pulses: (a) Gaussian, (b) spectrum 1, (c) spectrum 2.

In the validity range, as expected, the rms error is below 0.1 meaning a good measure quality. For most cases, the error is even below 0.02 meaning an excellent measure quality. When the spectrum shape is a bit complex, the error deterioration is due to the finite time gate and not intrinsically to the method.

The experimental implementation imperfections also degrade the measurement quality. In the following parts, the influences of optical setups, the non linear filter quality and the spectrometer are simulated to estimate their consequences on the SRSI method.

## 5    Effects of optical implementation setups imperfections

The setups sketched on fig.13, implement essentially two operations out of the non linear filtering: beam splitting/combining and delay generation. Different optical implementations of the SRSI method have been proposed [14, 15] (fig.13.a, b and c). Their main limitations are the dispersions introduced both on the pulse used for the XPW and on the replica. These two dispersions can be different. In the following we will distinguish the dispersion of the XPW channel $x_X$ and the dispersion on the pulse replica $x_S$.

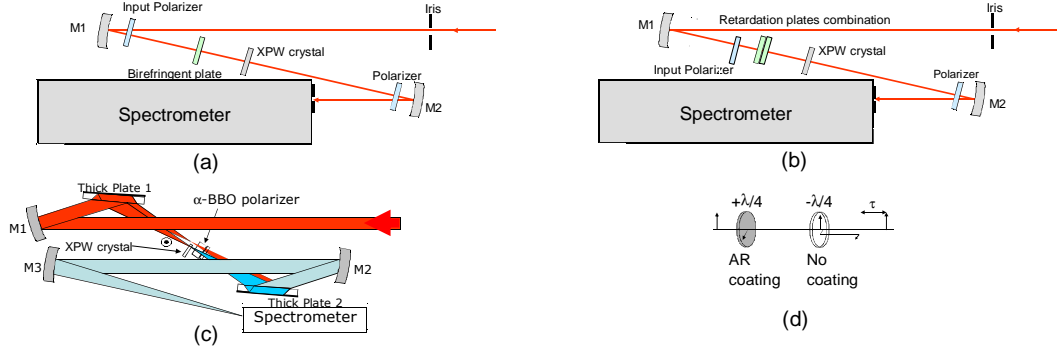

Fig.13: optical setups: (a) standard commercial system for 600-900nm pulses 25-120fs, (b) ultra-short pulse compatible setup 6-50fs and (c) ultra-wide band achromatic implementation, (d) waveplates combination for pulse replica generation.

The first setup (fig13.a) is collinear and bulky. It uses a birefringent plate for delay generation and beam combination. The input pulse is first filtered by a polarizer used in a double pass configuration. It perfectly defines its polarization direction. The precision of this polarization direction is essential for the XPW filter quality. After the polarizer, the pulse passes through a birefringent plate. At the output, two temporally delayed pulses aligned on the birefringent plate principal axes are obtained. Depending upon the orientation of the plate with respect to the input polarizer axis, the energy ratio between the two pulses can be adjusted. The first pulse (polarized orthogonally to the plane of the paper) is used to generate the XPW reference pulse in a $BaF_2$ crystal. As the XPW pulse polarization is orthogonal to the input one, a second polarizer used as analyzer selects only the XPW pulse. The second replica at the output of the birefringent plate propagates through the XPW crystal without any non linear effect. Its polarization is parallel to the XPW pulse. Finally, only the XPW and this second pulse will pass through the analyzer and interfere into the spectrometer.

As the SRSI method compares the spectral phases of a pulse replica and an XPW filtered one, the dispersion introduced on each pulses should be balanced. This setup introduces $105fs^2$ of dispersion on the pulse before the XPW generation. On the replica the dispersion introduced is on the order of $160fs^2$. It should be noted that the two pulses, the XPW generated and the pulse replica after the XPW crystal, are on the same polarization and in the same direction. As the SRSI measures the phase difference between these two pulses, any additional dispersion after the XPW crystal is balanced and thus has no influence on the measurement. It is therefore possible to use a bulk polarizer with thick material for the second polarizer or add any other optics.

The influence of the two dispersions can be included in the SRSI algorithm. We already define $x_X$ the relative rms chirp factor of the dispersion on the XPW beam path. It includes all the dispersion that the input pulse will face until it is filtered by XPW generation. By the same, $x_S$ is the relative rms chirp factor of the dispersion on the pulse replica beam path. For the setup of fig.13.a and for 10fs rms pulse duration, $x_x=1.05$ et $x_S=1.6$. The algorithm result can be expressed as:

$$x_\infty = \left(\left(4-\sqrt{16-9\left(\alpha\left(\frac{8(x+x_X)}{9+(\alpha(x+x_X))^2}\right)\right)^2}\right) \middle/ \left(\alpha^2\left(\frac{8(x+x_X)}{9+(\alpha(x+x_X))^2}\right)\right)\right) - x_X \quad (37).$$

As long as $|(x+x_X)| < \frac{3}{\alpha}$, $x_\infty = x$.

The SRSI method analysis of section 4 can be directly applied to $x+x_X$. Thus the only modification is a translation of the figures and validity range by $-x_X$. The dispersion on the pulse replica as far as it is known is completely removed in the measurement.

The limitation of this setup is thus only due to the translation of the validity range by the dispersion on the XPW beam path. For $105fs^2$, the range is still covering a reasonably large area around $x=0$ for pulse durations down to 9fs rms for a Gaussian shape. But for 5fs rms, $x=0$ is out of the validity range.

Conservatively this setup is specified for pulses down to 10fs rms i.e. about 25fs fwhm.

As only the XPW beam path dispersion influences the SRSI method, shorter pulses can be measured if less dispersion is introduced before the non linear generation. The setup of fig13.b is another implementation where the input polarizer is in single pass and is a thin film polarizer (250µm). The birefringent plate is replaced by a delay generator made from a combination of λ/4 waveplates (fig.13.d). The input pulse goes through a first λ/4 waveplate with axis rotated by an angle θ from the input polarization axis. This plate introduces a +λ/4 optical retardation. The pulse then goes through the second λ/4 waveplate whose axes are perpendicular to previous ones. The optical retardation introduced by this second waveplate is therefore - λ/4. The output pulse has seen no optical retardation and its polarization is the same as the input one. Let's consider know the back reflections. The first plate is anti-reflection coated. So there is no back reflection from this one. The second waveplate has no coating. A part of the pulse is reflected from the output face of this waveplate and re reflected from the input face in the same direction than the main pulse. This twice reflected pulse has gone once through the first waveplate (+λ/4), and three times through the second (-3λ/4). Globally it has been -λ/2 retarded. Its polarization is then rotated by 2θ. It is delayed from the main pulse by τ corresponding to twice the thickness of the second waveplate. For 100µm quartz plates, the delay generated is about 1ps while the dispersion on the main pulse used for XPW generation remains very small: 16fs$^2$. The delay can also be obtained from the reflections between the two plates' back sides. Its value is then not limited by the thickness of the plates. The delay value and dispersion introduced are independent.

Incorporating the thin film polarizer and the dispersion of the crystal, the total dispersion on the XPW beam path in this configuration is about 35fs$^2$. This allows measurement of pulses down to 3.5fs rms (8.2fs fwhm).

If this dispersion is pre compensated by a chirped mirror pair for example, then this limitation is completely removed.

The last setup (fig.13.c) is optimized to minimize dispersions and maximize the bandwidth. In the two previous setups, the input polarizer used is a thin film polarizer. Its bandwidth is limited to about 550-1100nm. The waveplates combination has about the same bandwidth limited range. Thus in this setup, the polarizer is replaced by a Fused silica Brewster polarizer (Thick plate 1). The input pulse reflected on the front side of this fused silica plate of thickness 6mm is perfectly polarized and has no dispersion on its beam path except the focusing mirror dispersion (<10fs$^2$). The bandwidth of such a polarizer is from 250nm up to 1200nm. The XPW pulse is then generated in the non linear crystal. An α-BBO Glan-laser polarizer is used to filter the XPW pulse. The extinction ratio is about 10$^4$ up to 10$^6$.

The global setup looks like a Jamin interferometer [16]. The two plates are equally thick plane parallel fused silica plates opaquely silvered or aluminized on the back surfaces. Their orientations are parallel. This interferometer is intrinsically balanced.

The XPW pulse is reflected on the back side of the second plate.

The pulse replica is generated from the back reflection of the first plate. It does not need to go through the non linear crystal. It goes through the α-BBO Glan-laser polarizer for dispersion balance. If the second plate has no coating on its front surface, then there is no reflection because of the Brewster angle incidence. Inserting an achromatic waveplate in this beam path to rotate the polarization can overcome this reflection problem. To keep the maximum bandwidth, the second plate has a low reflection coating on the first surface (Inconel coating bandwidth: >250-1200nm). The reflected pulse on this surface is automatically recombined with the XPW pulse. Their polarizations are parallel as they both go through the second polarizer.

Without any additional part, a delay between the two pulses is due to the XPW crystal. For ultra-large bandwidths, its thickness should be kept very small (200µm). Thus the delay is in the order of a picosecond. The delay can also be tuned by a small rotation of the second thick plate introducing a small different optical path between the two pulses.

This setup has no intrinsic limitation due to dispersion of optical components. But the efficiency of the Brewster polarizer is low with fused silica. It can be overcome by using a diamond plate for example. This setup is also fully compatible with a spatio-temporal SRSI characterization but this is out of the scope of this article.

This setup introduces no dispersion on XPW and its bandwidth is limited only by the material used for the Brewster polarizer (250-1200nm for Fused silica as example).

Thus the optical setup can be designed to introduce no bandwidth limitation and minimum pulse duration. But the SRSI measurement is not only limited by the optical setup.

## 6   Third order non linear filter limitations and defaults

The non linear temporal filtering effect is essential in this method. In the previous parts, it has been considered as ideal third order frequency conservative effect. In this part, the imperfections and limitations due to this non linear generation are introduced and discussed in the scope of the SRSI measurement precision.

Despite the fact that this method can be implemented by any non linear frequency conservative effect (Self-Diffraction for example [19]), this part will only deal with Cross-Polarized Wave generation. This effect has been particularly studied by A.Jullien, O.Albert, N.Minkovski and S.M.Saltiel [17-20] and through the thesis of L.Canova [21].

This effect is based on the anisotropy of the tensor $\chi^{(3)}$. For cubic crystal (isotropic), this non linear anisotropy induces a coupled wave generation between the input wave and an orthogonally polarized wave. This cross-polarized wave can be expressed [20] $E_{XPW}^+$ as:

$$\frac{dE_{XPW}^+}{dz} = -i\gamma_\perp |E^+|^2 E^+, \frac{dE^+}{dz} = i\gamma_{//} |E^+|^2 E^+ \quad (36) \text{ where } |E_{XPW}^+| \ll |E^+|, \gamma_{//} = \gamma_0 \left[1 - (\sigma/2)\sin^2(2\beta)\right]$$

and $\gamma_\perp = -\gamma_0 (\sigma/4)\sin(4\beta)$, with $\gamma_0 = (6\pi/8\lambda n)\chi_{xxxx}^{(3)}$ and $\sigma = \left[\chi_{xxxx}^{(3)} - 2\chi_{xyyx}^{(3)} - \chi_{xxyy}^{(3)}\right]/\chi_{xxxx}^{(3)}$ the anisotropy of the $\chi^{(3)}$ tensor. Angle β is the angle between the input polarization and the [001] axis.

These relations neglect the depletion of the input wave, self-phase modulation of XPW and the possible effects of cross-phase modulation. We neglect the imaginary part of the coupling constants ($\gamma_\perp = \text{Re}[\gamma_\perp]$ and $\gamma_{//} = \text{Re}[\gamma_{//}]$). These hypotheses stand for low efficiency XPW generation. The efficiency can be expressed as:

$$\eta = \left|\frac{E_{XPW}^+}{E^+}\right|^2 = 4\left(\frac{\gamma_\perp}{\gamma_{//}}\right)^2 \sin^2\left(\gamma_{//} E_0^2 L/2\right) \quad (37).$$

For low intensities, it simplifies to:

$\eta = \gamma_\perp^2 E_0^4 L^2$ (38), for a plane monochromatic wave.

The efficiency is dependant on the power of the input pulse. This non linear effect cannot be used at very low power. It can be tuned by the length of the crystal as long as the plane wave approximation is still valid i.e. the Rayleigh length at the focus of a Gaussian beam. There is no phase matching condition. This non linear effect is intrinsically achromatic. Its dependence on the crystal orientation is not stringent.

For large bandwidth Gaussian pulses, an optical frequency factor is expected as the waist of the focused beam depends upon the wavelength ($\omega^2$) and as the length of the crystal expressed in wavelength varies ($L^2 \to \omega^2$). The XPW generated should have the form:

$\left|E_{XPW}^+\right|^2 \propto \omega^4 \gamma_\perp^2 \left|E^+\right|^6 L^2$ (39), without any assumption on the optical frequency dependence of the coupling constant.

As suggested by Miller's rule, the non-linear order should change with the wavelength as the linear susceptibility:

$\chi^{(3)} \propto \left(\chi^{(1)}\right)^4$ (40) where $\chi^{(1)} = \left(n^2 - 1\right)/4\pi$.

The XPW crystals considered (LiF and BaF$_2$) have very low dispersion of the refractive index in the spectral bandwidth of interest (except in UV range). The first order susceptibility can be considered as nearly independent upon the optical frequency. The coupling constant is then also independent of the frequency.

By these rules of thumb, the spectrum intensity of the XPW should have a frequency dependence as:

$$\left|E_{XPW}^{+}\right|^{2} \propto \omega^{4} \gamma_{\perp 0}^{2} \left|E^{+}\right|^{6} L^{2} \quad (41).$$

Thus for the simulation of the XPW in the SRSI, a frequency factor $\omega^4$ should be introduced as checked experimentally (blue shift of the XPW pulse). This approximate rule explains the qualitative use of the comparison between XPW simulated and measured spectra in the validity criteria. It has no consequence on the spectral phase filtering because the temporal filter width is not significantly modified.

The XPW efficiency is limited to few percent. At this level, white light generation in the crystal may happen. For the measurement, it is essential to avoid any distortion and parasitic effect, so the efficiency that is considered is in between few per thousand and a percent.

This low efficiency strongly enhances the requirement on the polarizers extinction ratio. The extinction should keep a spurious pulse associated with the XPW at a level of one over thousands. Combined with the efficiency, this means that the polarizers should have extinction ratio at least $10^4$. The quality of optics in between the polarizers has to be good enough to not deteriorate this extinction ratio.

This spurious pulse influences the measurement dynamic. The evolution of the rms error profile in the validity range versus this extinction ratio is illustrated on fig.14. The error is negligible for extinction ratio lower than $10^5$ (fig14.a and b), low for $10^3$, but increases significantly for $10^2$ (fig.14.d). In any case and even for lower ratios, the error is still nullified for flat phase pulses. In the SRSI method, this spurious pulse lowers the estimation of the phase because it adds to the reference pulse a pulse identical to the input pulse phase (without differential setup dispersion considered). At the limit, when the XPW pulse is negligible, the spectral phase measured is null because it is the difference of the input pulse with its own replica. The consequence is a diminution of the validity range. But still on this narrower validity range, the phase determination is accurate ($\varepsilon<0.02$).

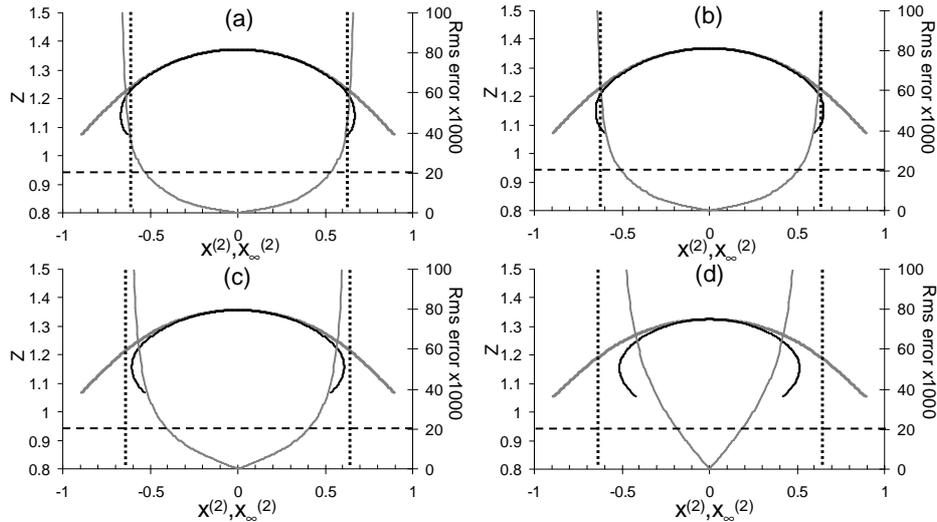

Fig.14 : XPW parasitic pulse influence on the Z and rms error versus x and $x_\infty$: (a) ratio XPW pulse energy over parasitic pulse equal $10^6$, (b) $10^4$, (c) $10^3$, (d) $10^2$.

In feedback loop compression optimization with a pulse shaper, this effect is similar to an attenuation of the feedback. It only decreases the convergence rate.

The last important effect in the XPW generation comes from the dispersion of the crystal itself [22]. For short pulses, this dispersion has to be low enough to not stretch the pulses and thus loose its temporal filtering characteristics. Using thin crystal (200μm LiF), the dispersion is below 4fs$^2$ at 800nm but at the expanse of the efficiency again. One part of the dispersion can be taken into account in the SRSI algorithm but as long as the temporal filtering effect is still efficient.

A balance between dispersion, efficiency, white light generation (or other non linear spurious effect), polarizers extinction ratio has to be found.

## 7 Spectrometers imperfections consequences

In the SRSI method, the experimental signal is measured by a spectrometer with a CCD linear detector. Spectrometers characteristics and imperfections that clearly impinge on the measurement are studied by the estimation of the rms error: noise level and dynamic of the spectrum measurement, digitization, non linearity of the detector, pixelization, bandwidth, resolution and optical setup (spectrograph) defaults.

### 7.1 Noise and dynamic

The noise and dynamic of the spectrometer come from the CCD linear detector and its digitization electronic circuit. The dominant noise in our case is an offset white noise due to both dark current in CCD pixel and Johnson type electronic noise of the readout and analog-to-digital conversion.

It has to be noted that the photon noise proportional to the square root of the signal level is not relevant in our case. This point is beyond the scope of this article and is partially addressed by Jacubowiez and al. in [22] and in Appendix C.

The noise level can be estimated by the dynamic of the CCD detector commonly expressed as the ratio of the maximum signal over the noise. This dynamic is in the range of 300 to 20000 for commercial systems.

One should pay attention that the offset due to this noise is usually hidden by signal processing of the spectrogram directly on the electronic readout circuit or by software.

The use of a bi dimensional CCD detector can enhance the dynamic as it reduces the noise by averaging as the square root of the number of lines taken into account.

### 7.2 Digitization

The analog-to-digital conversion of the signal may also limit the dynamic because of digitization of the signal at 8 to 16 bits. With 16bits no modification on the rms error can be seen on computer simulation. At 8bits, the rms error is raised but still at level below 0.02.

### 7.3 Non linearity of the detector

CCD detectors are more linear than CMOS ones. But used on the complete dynamic range, saturation appears at high level signals. This saturation can be simulated as a non linear response of the detector to the signal (fig.15.a). It modifies the spectrum intensity. As the SRSI signal is an interferogram (fig.15.b), it creates pulse replica in the time domain: a kind of "temporal harmonics". Thus its level can be estimated directly on the Fourier transform module of the spectrogram (fig.15.c). Also the main part of this effect is filtered out in the FTSI data processing by the digital filter in the time domain. Except when its distortion is too important and affects directly the intensity spectra of the pulses, it is completely filtered out (fig.15.c).

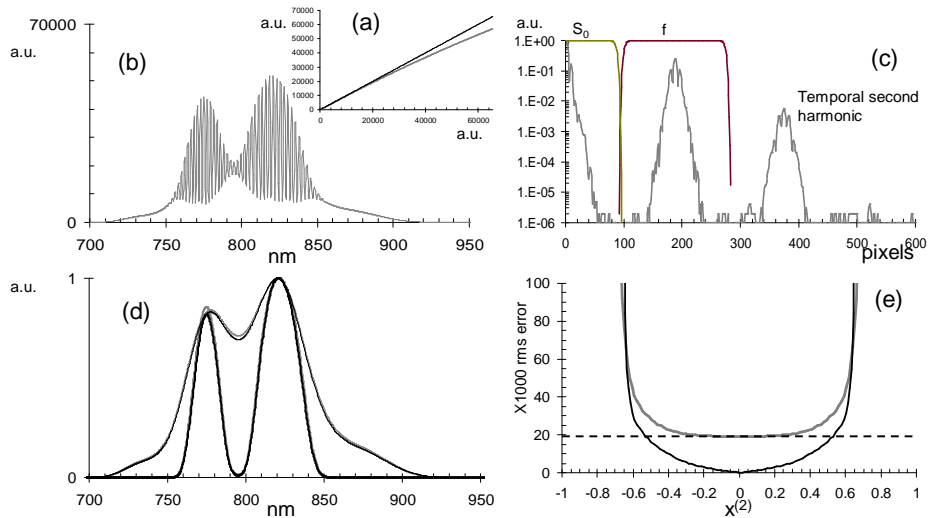

Fig.15: saturation of the CCD detector: (a) measured signal versus input signal, (b) interferogram with saturation, (c) its Fourier transform module with pulse replica, (d) spectra reconstruction and (e) rms error versus $x^{(2)}$ with no saturation(black) and saturation(grey).

Even with a saturation limited about 15%, the rms error is still close or below 0.02 in the validity range (fig15.e). The measurement remains very good.

The saturation can be annealed by the data processing using either an initial calibration or the pulse replica.

### 7.4 Pixelization

The CCD detector is pixelized. This pixelization is a sampling issue of the signal. Aliasing and defects explained by the sampling Nyquist-Shannon theorem appear. In the computer simulation used, the interferogram is sample on the number of pixel. The pixelization is then directly included in this model. Its consequences depend upon parameters more relevant from the optical setup (delay) and the data processing. It is illustrated in the section dedicated to the signal processing hereafter.

### 7.5 Spectrometer optical setup defaults

The spectrometer optical setup is based on a Czerny-Turner design as shown on fig.16, introducing the notation used herein.

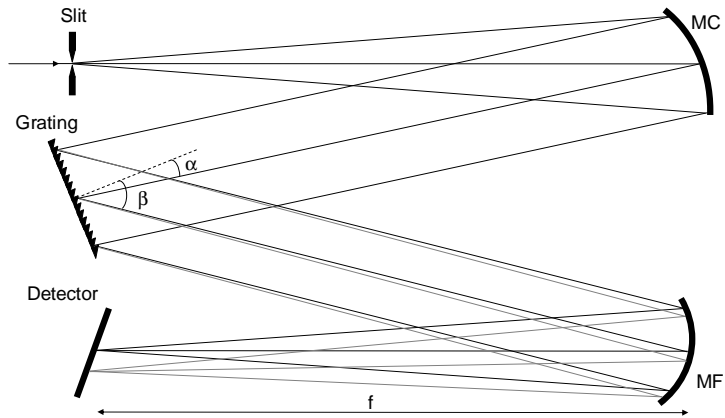

Fig.16: Classical Czerny-Turner spectrometer design: MC, spherical collimating mirror, MF, spherical focusing mirror. f effective focal length, grating with angles of incidence $\alpha$ and diffraction $\beta$.

A divergent wavefront from the entrance slit is collimated by spherical mirror MC and diffracted in the tangential plane by the grating. The light is then focused by the spherical mirror MF onto the detector. Extensive literature on this configuration design exists. One can refer for example to imaging configuration optimizations [23-25].

Even optimized, this optical setup defines the basic characteristics of the spectrometer as bandwidth, resolution and spectral response.

#### 7.5.1 Grating equation

For the data processing and in particular the Fourier transform, It is greatly simpler if the signal is regularly sampled in frequency. Unfortunately, the grating used to disperse the wavelength angularly follows an equation in wavelength: $\sin \beta = \sin \alpha + \lambda/D_G$ (38) for the first order of diffraction with $D_G$ groove density of the grating commonly expressed in grooves/mm. The relation of spatial dimension in the detector plane with the wavelength expressed hereafter in the resolution section is pseudo-linear in wavelength: $x \propto \lambda - \lambda_0$. This classical problem of re sampling in Fourier Transform Spectral Interferometry is also present in Spectral Phase for Interferometric Direct Electric-field Reconstruction. It is solved by fine calibration of the spectrometer as demonstrated by Dorrer [26].

#### 7.5.2 Optical aberrations

Optical imaging system of the spectrometer is often based on Czerny-Turner configuration to optimize the aberrations (fig.16). These aberrations lead to a modification of the Point Spread Function in the imaging field i.e. on the CCD detector. The consequence is that the PSF depends upon the wavelength and has not a constant profile. It impacts the resolution of the spectrometer as detail hereafter.

*7.5.3 Non constant transmission*

The measured signal is weighted by spectral amplitude due to the optical components transmissions or reflections and the wavelength dependant sensitivity of the CCD detector. It can be corrected by an initial calibration of the spectrometer using a calibrated white light.

*7.6 **Bandwidth and resolution***

The bandwidth and resolution of the spectrometer have major importance on the temporal resolution and measurement range.

The bandwidth expressed in pulsation as $\Delta\omega$ defines directly the temporal resolution as: $\delta t = 2/\Delta\omega$. The bandwidth of the spectrometer is necessarily larger than the pulse bandwidth. This temporal resolution can be artificially enhanced by zero padding in the frequency domain.

The resolution $\delta\omega$ by inverse determines the maximum temporal range that can be used for the SRSI: $\Delta T_{spectro} = 2/\delta\omega$ (39). As for FTSI, the digital filtering in time domain limits the maximum range to one third of this maximum temporal range: $\Delta T_{SRSI} < 2/3\delta\omega$ (40).

The resolution in a CCD spectrometer is a combination of the pixel size, the optical setup imaging quality and the dispersion of the grating. The linear dispersion of the spectrograph defines the extent to which a spectral interval is spread out across the focal field of a spectrometer [24] and is expressed in nm/mm. For the diffracted beam at a central wavelength, it is given by:

$$\frac{d\lambda_0}{dx} = \frac{\cos\beta_0}{D_G f}$$ (41) where f is the effective focal length, $D_G$ the groove density of the grating (grooves/mm), $\beta_0$ is the angle of diffraction at $\lambda_0$.

The linear dispersion for any wavelength other than that wavelength is modified by the cosine of the angle of inclination $\gamma$ at wavelength $\lambda$:

$$\frac{d\lambda}{dx} = \frac{\cos\beta_0 \cos^2\gamma}{D_G f}$$ (42) where $\gamma = \beta(\lambda) - \beta_0 \approx \frac{D_G}{\cos\beta_0}(\lambda-\lambda_0) + \beta_0$.

The relations between the spatial dimension on the spectrometer and the wavelength are given by:

$$x(\lambda) \approx \frac{D_G f}{\cos\beta_0}\left(1-2\beta_0^2\right)(\lambda-\lambda_0) - \frac{4D_G^2 f \beta_0}{\cos^2\beta_0}(\lambda-\lambda_0)^2 - \frac{2D_G^3 f}{\cos^3\beta_0}(\lambda-\lambda_0)^3$$ (43),

$$\lambda(x) \approx \lambda_0 + \frac{\cos\beta_0}{D_G f}\left(1+2\beta_0^2\right)x + \frac{4\beta_0\cos\beta_0\left(1+2\beta_0^2\right)}{D_G f^2}x^2 + \frac{2\cos\beta_0\left(1+2\beta_0^2\right)^2}{D_G f^3}x^3$$ (44).

It follows that the resolution given by the pixel size is not constant over the CCD detector even in wavelength. To minimize this effect, one should choose a high density groove and a long focal length. With standard USB spectrometers ( Avantes, 75mm focal length and 600 lines/mm groove density), the coefficient are :

$$\lambda(x) \approx \lambda_0 + 2\cdot 10^{-5} x - 10^{-12} x^2 - 10^{-21} x^3$$ (45).

Thus at the first order: $\delta\lambda(x) \approx \frac{\cos\beta_0}{D_G f}\left(1+2\beta_0^2\right)\delta x_{pixel}$ implies in frequency near the central wavelength $\lambda_0$:

$$\delta\omega)_{\lambda_0} \approx \frac{1}{\lambda_0^2}\frac{\cos\beta_0}{D_G f}\left(1+2\beta_0^2\right)\delta x_{pixel} \approx 0.4THz$$ (46).

The maximum temporal range of the SRSI is then: $\Delta T_{SRSI} \leq 2/3\delta\omega \approx 1.5\,ps$.

As seen from $\lambda(x)$, the resolution of the spectrometer is more complex because on large bandwidth, it cannot be considered as constant.

Diffraction limit of the imaging system and aberrations should also be added with this pixel resolution to get the instrumental line profile or Point Spread Function (PSF) of the spectrometer.

The diffraction limit comes from the aperture of the optical setup expressed by the numerical aperture NA. For 1 dimensional spectrometer, it has the form of a cardinal sinus function at one wavelength $\lambda$:

$$PSF_{diff}(\omega) = \sin^2 c\left(\pi \frac{Lx(\lambda)}{\lambda f}\right)$$ (47) where $x(\lambda)$ is the position on the CCD detector.

The measured signal is the correlation of the input signal with this function. Thus in the temporal domain, the amplitude is limited by its Fourier transform as envelope.

Without aberrations and by considering that locally on a small bandwidth around the central wavelength $\lambda_0$ $x(\lambda) = x_0$, the temporal envelope has the form of a triangle (Fourier transform of sinc$^2$) with base $L(N_{pixel}\delta x_{pixel})/2cf \approx 6.7\,ps$ with $L/2f$ the numerical aperture of the optical setup typically 1/8 and 0.07 in our example, CCD width $N_{pixel}\delta x_{pixel} = 28.7\,mm$ ($\approx 14\mu m \cdot 2048$) and in the middle of the bandwidth $x_0 = N_{pixel}\delta x_{pixel}/2 = 10\,mm$.

If we consider larger bandwidth then the approximation $x(\lambda) = x_0$ cannot strictly be done. The PSF is more complex and frequency dependant:

$$PSF_{diff}(\omega) \approx \sin^2 c\left(\frac{L}{2cf}\omega\left(x_0 + \frac{2\pi c A_1}{\omega} + \frac{(2\pi c)^2 A_2}{\omega^2} + ...\right)\right)$$ (48).

Its Fourier transform cannot be expressed analytically. The temporal envelope of the signal is wavelength dependant. As it represents a spectro-temporal variation, it cannot be corrected through an appropriate filter in the time domain only. This subtle effect becomes significant only if the frequency variation is weighty. In the following it will be neglected as $x_0 \gg \frac{2\pi c A_1}{\omega} \gg \frac{(2\pi c)^2 A_2}{\omega^2}$.

*7.7 Synthesis on spectrometer imperfections on SRSI method*

The specifications of the spectrometer determine the dynamic of the measurement, the resolution and range in wavelength and time. The significant effects listed here before can be compensated by calibration or by an appropriate signal processing as described in the following section.

## 8 Data processing

The data processing of the SRSI is schematized on fig.3. It consists in six principal steps. The interferogram is pre processed. An inverse Fast Fourier Transform is used to switch into the temporal domain. The temporal signal is then digitally filtered and processed. A Fast Fourier Transform swaps the signal back into the spectral domain. The SRSI algorithm and algebraic spectra calculation are applied to recover the spectral signals. Additional spectral digital filters and inverse Fast Fourier Transform is used to obtain the temporal results: intensity and phase.

*8.1 Interferogram pre processing*

The measured interferogram is 16 bits quantized output of the spectrometer. It is converted into double precision floating point numbers.

It can be filtered to restrict the bandwidth to the useful spectral range.

In case of saturation of the signal, if the saturation curve is calibrated, it can be corrected directly on the signal by inversion. A more complex algorithm based on the "temporal second harmonic" elimination can also be implemented in here without calibration.

The spectral signal is also re sampled on a regular frequency comb by an interpolation using zero padding in the dual space before spline interpolation.

*8.2  Fast Fourier transform and Quantization noise power estimation*

The initial 16 bits 1D interferogram has a dynamic of $20\log_{10}(2^{16}) \approx 96 dB$ (49).

The signal pre processed is Fast Fourier transform. In fact it has been already inverse Fast Fourier Transform and Fourier transform for the re sampling process. The quantization noise introduced in FFT is difficult to estimate. Weinstein (1969) showed theoretically that the signal-to-noise ratio (SNR) is proportional to 1/i for N=$2^i$ points for floating-point arithmetic [27]: $SNR_{result} = \frac{|X(n)|^2}{\sigma(n)^2} \approx \frac{2^{2p}}{0.42 \cdot i} > 600 dB$ (50) where p=52 is for double precision floating point numbers.

For any digital filter having finite impulse response (FIR filters), the SNR for double precision floating point is even smaller than 600dB.

So the quantization noise introduced in the data processing can be neglected.

The more limiting digital operation, the spline interpolation used to re sampled the signal on a regular frequency comb, has an estimated SNR limit about 240dB.

The digital dynamic is thus limited to about 100dB by 16 bits quantization of the spectrometer.

*8.3  Temporal digital filters*

The data are now in the temporal domain. Two digital filters are used to isolate $f(t)$ and $S_0(t)$. In this domain, the spectrometer PSF (temporal triangular envelope) can be de convolved by multiplying by the inverse of the envelope. In this case the two temporal filters are not flat but their shapes compensate for this envelope.

One important issue of the measurement happens in this processing: the temporal aliasing.

The temporal range of the SRSI is defined by two delays: the delay used for the SRSI $\tau_{SRSI}$ and the two temporal filters widths separate the temporal domain into three sub parts (two in the positive time domain). The two temporal filters have the same width $\Delta T_{SRSI}$. This width has a maximum value defined by:

$$\Delta T_{SRSI} \leq \tau_{SRSI} \leq (\Delta T_{spectro} - \Delta T_{SRSI})/2 \quad (51).$$

This double inequality expresses the limits of both delays $\tau_{SRSI}$ and $\Delta T_{SRSI}$ in function of the spectrometer resolution $\Delta T_{spectro} = 2/\delta\omega$.

Let's consider a SRSI delay of $\tau_{SRSI} = 2 ps$, a pulse added with a post pulse delayed by $\tau = 1.7 ps$ (fig.17.a for the spectral intensity and b for its temporal positive time representation). Then the combination of pulses in the SRSI corresponds to four pulses in the temporal domain (fig.17.d). The pulse and one of the replicas are then separated by only $\delta t = \tau_{SRSI} - \tau = -300 fs$ corresponding to the SRSI delay minus the post pulse delay. The spectra is then exhibiting modulation totally different compared to the initial one (fig.17.c).

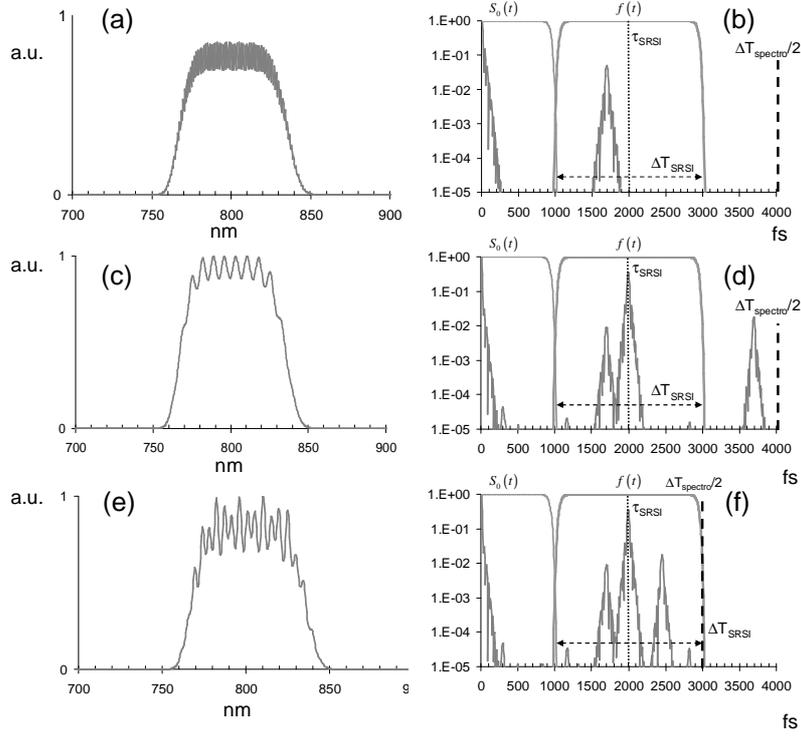

Fig.17; illustration of temporal aliasing effect: (a) initial pulse spectral intensity, (b) amplitude in temporal domain, (c) SRSI reconstructed spectral intensity from (d) SRSI temporal amplitude signal of this input pulses with 2ps delay pulse replica in temporal delay, (e) SRSI reconstructed spectral intensity from (f) SRSI temporal amplitude signal with aliasing from SRSI delay and spectrometer (the resolution of the spectrometer is different in this case).

This effect can be combined with spectrometer aliasing (fig.17.e, f) if $\left|\delta t = \Delta T_{spectro} - \tau - 2\tau_{SRSI}\right| \leq \Delta T_{SRSI}/2$ to give an even more complex spectral shape. This aliasing is inherent to sampling issue in the time domain due to the spectrometer resolution and the temporal filter used to isolate the DC and AC terms.

The comparison with the initial spectral intensity points out this effect and can be used to unfold the replica.

### *8.4 SRSI phase algorithm and spectra algebraic calculation*

The SRSI algorithm has been described extensively in a previous section. This part focuses onto quantization noise and data processing limitations of the algorithm.

The phase iterative algorithm uses two Fourier transforms and an XPW simulation effect. The FFT quantization noise is negligible. The XPW simulation uses simple multiplication and should not introduce any significant digital noise.

For the spectral amplitude estimation it uses square roots of linear combinations of $f(t)$ and $S_0(t)$. These linear combinations have to be positive onto the complete spectral bandwidth. This corresponds to an XPW signal higher than the pulse replica one. Within this condition, no significant noise or mistake appears in the processing. The input spectrum is even extracted from this noise because it is only due to the interference pattern. As the XPW spectrum is larger, this interferometric part is upper than the noise level even on the side of the spectrum. The dynamic measurement of the input spectrum is enhanced compare to the spectrometer dynamic by a kind of self-heterodyne detection (c.f. appendix C).

From this point the spectral results are completed: input pulse spectral amplitude and phase are recovered, XPW ones also.

*8.5  Spectral filtering and inverse Fourier transform: back in the time domain*

The most interesting parameter of such ultrafast lasers is in general the temporal intensity. An inverse FFT of the previous spectral results is sufficient to get the temporal intensity and phase. Experimentally, it is of interest to filter one part of the spectral noise before. Digitally, the noise is low enough to not modify significantly the signal. Nevertheless, the filter has to be carefully set to avoid any bandwidth shortening of the real signal.

## 9  SRSI limits

The SRSI limits can now be determined and computer simulated from the previous part. This section highlights key parameters to optimize for some relevant examples: maximum, minimum pulse durations and temporal dynamic.

*9.1  Maximum pulse duration*

One can ask the limit of pulse duration that can be measured with SRSI method. There can be two kind of long duration pulses: bandwidth limited pulse and spectral phase stretched pulse.

The spectral phase limit has been reviewed in the section dedicated to the validity range of the measurement. All the parameters are defined in statistics meaning. The most stringent spectral phase order is the chirp. The limit is given by:

$$\left| x^{(n)} \right| \leq X_{\lim} = B/\alpha_n = \left((n-1)! B\right) \Big/ \left((\alpha)^{n-1} \sqrt{\beta_{2(n-1)}}\right)$$ (52) where n=2 for chirp.

In this case, the limit is about twice the duration of the Fourier limited pulse.

So one should ask the limit for Fourier limited pulse. To measure a long pulse, the delay of the pulse replica has to be long enough to avoid any covering between the XPW pulse and the replica. As shown on fig.17 with the aliasing examples, the spectrometer resolution has to be small enough to also avoid any aliasing effect and covering. So for estimated rms duration $\Delta\tau$ of the pulse, the SRSI delay should be $\tau_{SRSI} \approx 3\Delta\tau$ and the resolution $\delta\omega$ given by:

$$\delta\omega \leq 2/3\tau_{SRSI} \approx 2/9\Delta\tau$$ (53).

Another important limitation is the ability to effectively generate XPW. This limit is counter balanced if the pulse energy can be tuned. Longer pulse measurement requires higher energy per pulse. The typical order is about 1µJ for 10fs pulse, and it scales as: $E_{pulse} = \left(\dfrac{1\mu J}{10 fs}\right) \Delta\tau$. For example, 1ps duration pulse corresponds to a pulse energy of 100 µJ. This can be decreased by using longer XPW crystals. It scales as the square of the crystal length. So the relation is: $E_{pulse} = \left(\dfrac{1\mu J}{10 fs}\right)\left(\dfrac{L_{XPW}}{1mm}\right)^2 \Delta\tau$. By using a 4mm crystal length instead of 1mm, the 1ps pulse can be measured with about 6µJ. The crystal length is limited by the Rayleigh length depending upon the characteristics of the input beam and the optical setup numerical aperture. In SRSI standard configuration, this length is about 4mm for a waist of 25µm resulting of the focus of a 1mm beam through a 100mm lens. Longer crystal requires a modification of the optical setup.

Thus for long pulse measurement of rms duration $\Delta\tau$, the key parameters are:

- the resolution of the spectrometer: $\delta\omega \leq 2/9\Delta\tau$,
- the SRSI delay: $\tau_{SRSI} \geq 3\Delta\tau$,
- the energy per pulse for a XPW crystal length: $E_{pulse} = \left(\dfrac{1\mu J}{10 fs}\right)\left(\dfrac{L_{XPW}}{1mm}\right)^2 \Delta\tau$.

Pulses up to about a picosecond can therefore be measured if their energy per pulse is in the range 10 to 100µJ.

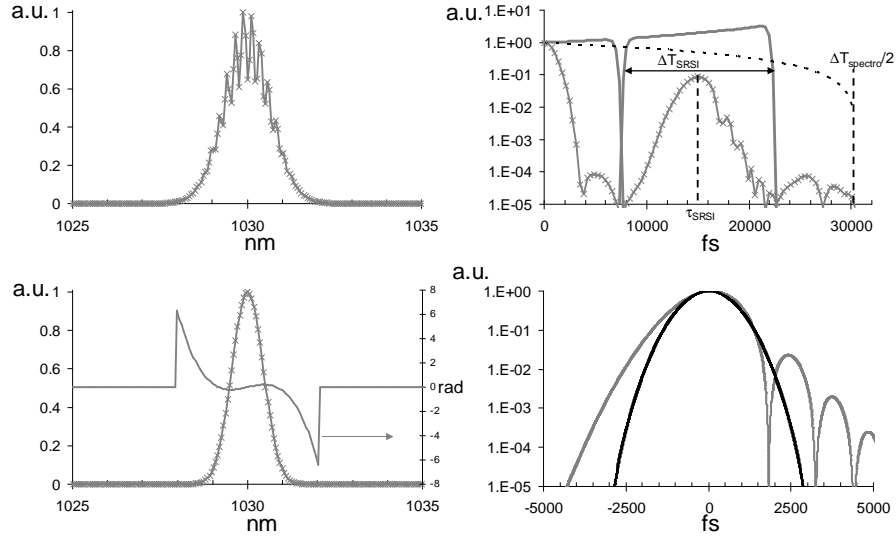

Fig.18: 0.8 ps rms (1.6ps fwhm) pulse measurement (ε=0.28): (a) initial SRSI spectrum with resolution 0.06nm, (b) temporal filters with replica delay (15ps), digital filters (thick grey) and temporal envelope(dotted black), (c) measured spectral phase and intensity, (d) temporal intensities of measured pulses (grey) and its Fourier transform limit (black).

The quality of the measurement is mainly limited by the temporal limited excursion that cut the pulse wings. For close to Fourier transform pulse, the rms error decreases down to ε=0.002.

### 9.2 Minimum pulse duration

Few cycles pulses are of interest for many emerging and developing ultrashort laser fields. Their measurement is still a hard task. As pointed out through the relative chirp factor $x^{(2)}$, pulse duration below 5fs rms are distorted in time by even small dispersions. $25fs^2$ more than double the pulse duration. This corresponds to about 1.2m of propagation in air or 0.5mm of BK7. The other relevant characteristic is the very large bandwidth associated.

The limitations of the SRSI for such ultrashort pulses are thus the dispersion of the optical components and the bandwidth limitations either from the spectrometer or the non linear effect. Hopefully, the XPW is achromatic and thus has no intrinsic limitation of bandwidth.

For these ultrashort pulses, one should consider the Ultra Wide Band setup (fig.13.c). The dispersion introduced on both beams is less than $15fs^2$: less than $10fs^2$ from the reflective optics, $5fs^2$ from the thin XPW crystal.

The most stringent feature is the XPW spectral bandwidth. Any spectral cut can affect the measurement. If the XPW spectrum is cut by the spectrometer then in the temporal domain, $S_0$ and f are mixed together. The dynamic and quality of measurement are then strongly affected. It is necessary to window the global SRSI initial spectrum to avoid such effect. An adequate window that will not limit to much the bandwidth is a cosine tapered window. This window is defined as:

$$y = \begin{cases} 0.5x_i\left(1-\cos(2\pi i/2m)\right) & \text{where } i = 0,1,...,m-1 \\ x_i & \text{where } i = m,...,n-m-1 \\ 0.5x_i\left(1-\cos(2\pi(n-1-i)/2m)\right) & \text{where } i = n-m,...,n-1 \end{cases} \quad (54),$$

where $m = [nr/2]$, with n the number of element in $x_i$ and r the ratio of the total length of the tapered section to the whole signal length. If $r \leq 0$, the window is equivalent to a rectangular window. If $r \geq 1$, the window is equivalent to a Hanning window. For SRSI, $0.05 \leq r \leq 0.2$.

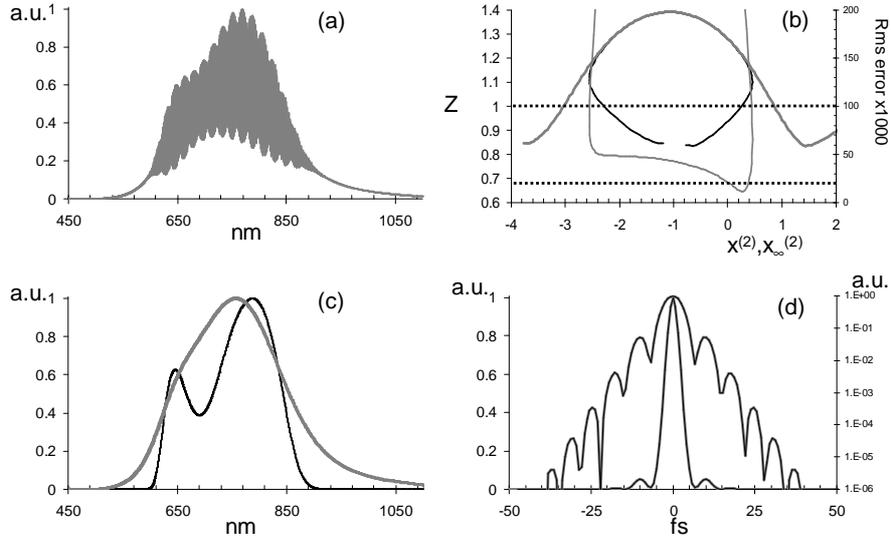

Fig.19: ultrashort pulse example with a super-gaussian spectral shape: (a) initial SRSI spectrum, (b) Z versus x and $x_\infty$, rms error versus x, (c) input and XPW measured spectral intensities, (d) temporal intensities of measured pulses (grey) and its Fourier transform limit (black).

The input pulse considered is an asymmetric "super Gaussian" of order 3 with a Gaussian hole (spectrum 3): $\left|\tilde{E}^+(\omega)\right| = \left(1 - H_{depth} e^{-\frac{1}{2}\left(\frac{\omega - \omega_{Hole}}{\Delta\omega_{Hrms}}\right)^2}\right) e^{-\frac{1}{2}\left(\frac{\omega - \omega_0}{\Delta\omega_{rms}}\right)^6}$, where $H_{depth} = 0.4$ is the hole depth, $\Delta\omega_{Hrms} = 2\pi c \left(100/800^2\right) nm^{-1}$ is the hole width, $\omega_{Hole} = 2\pi c / 690nm$ is the central position of the hole, $\Delta\omega_{rms} = 2\pi c \left(240/800^2\right) nm^{-1}$ the width and $\omega_0 = 2\pi c / 720nm$ the central pulsation of the super Gaussian part. Its shape is similar to spectrum 2 used in previous section but with larger bandwidths. Its Fourier transform limit has 3.6fs rms and 6.1fs fwhm durations.

The quality of measurement is good on $x_0^{(2)}$ range covering -2.25 to 0.2. Close to $x_0^{(2)}$=0 the measurement is even excellent. But if the input pulse is chirped positively $x_0^{(2)}$>0.25, the measurement is out of range.

This shift is due to the dispersion even limited to 15fs². Minimizing even further the dispersion decreases this shift. One should note that this dispersion can be pre compensated by using chirp mirrors for example. Then the limit comes from the XPW generation and the spectrometer bandwidth.

The XPW crystal dispersion can be decrease by using thinner crystal. But as the efficiency is proportional to the square of this thickness, a tradeoff in between pulse energy, XPW efficiency, polarizer extinction ratio and crystal dispersion has to be made. One can consider 5fs² as the ultimate low dispersion limit. The shift is then partially annealed and the measurement is good over -1.9 to 1 relative chirp.

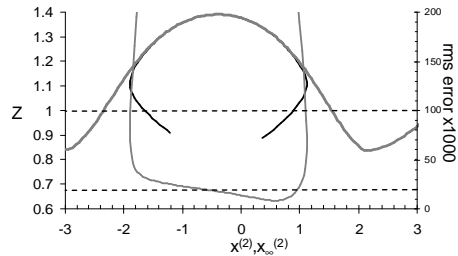

Fig.20: Z versus x and $x_\infty$, rms error versus x for dispersion limited to 5fs².

The spectrometer bandwidth limitation is essentially due to the imperfections described in here before. In the computer simulations of this section, it has not been taken into account. It influences more specifically the temporal dynamic.

### 9.3 *Temporal dynamic limitation*

The temporal dynamic is estimated as the maximum temporal intensity over the noise. Its limit is a combination of optical setup extinction ratio, spectrometer dynamic and imperfections, data processing. To illustrate the ability of the SRSI method to measure pulses over a good temporal dynamic, one should consider two pulses examples (spectrum 2 and spectrum 3). A parasitic pulse identical to the main pulse with a delay of 300fs is added. Its relative level will be adjusted to be one decade over the noise of the lower dynamic i.e. $10^{-4}$. Two different SRSI setups are considered. One corresponds to a standard device: USB spectrometer with 300:1 dynamic, optical setup extinction $10^4$. The other considers top quality components, extinction ratio $10^6$, spectrometer dynamic 10000:1 with averaging over 100 lines without significant aberrations.

As pointed out in [28], the self-heterodyne nature of the SRSI measurement combined with the compression ratio due to its measure in the time domain, gives to this method an improve in temporal measurement contrast limitation given by the rule of thumb:

$C_t = F \cdot N_{pixels} \cdot SNR_\omega$ (59) where $F = \sum |\tilde{E}_i^+| / N_{pixels}$ F is the filling factor, $N_{pixels}$ is the number of pixels and $SNR_\omega$ is the spectral signal to noise ratio i.e. the spectrometer dynamic. This rule of thumb represents the temporal dynamic gain due to the compression in time domain. The measure is done on $FN_{pixels}$ in the spectral domain but corresponds to a few pixels pulse in time domain.

In details (c.f. Appendices B and C), the contrast is also enhanced by noise reduction on the input signal spectrum. The rule of thumb is thus a conservative rule for contrast.

#### 9.3.1 *Standard SRSI device dynamic*

The SRSI setup combines a 300:1 dynamic, 16 bits digitization, 0.25nm resolution spectrometer with a dispersion compensated UltraShort Pulse optical part (fig.10.b). The parameters used in this simulation are:
- thermal noise 0.33%=1:300,
- 16 bits analog-to-digital conversion,
- 550-1050nm spectral bandwidth with 2048 pixels,
- Extinction ratio for the polarizer compared to XPW pulse: $10^4$.

The rule of thumb temporal contrast estimations are $C_t \approx 10^5 = 50 dB$ for spectrum 2 and $C_t \approx 5 \, 10^5 = 57 dB$ for spectrum 3.

The other parameters have no effect on the temporal dynamic or can be corrected through the data processing.

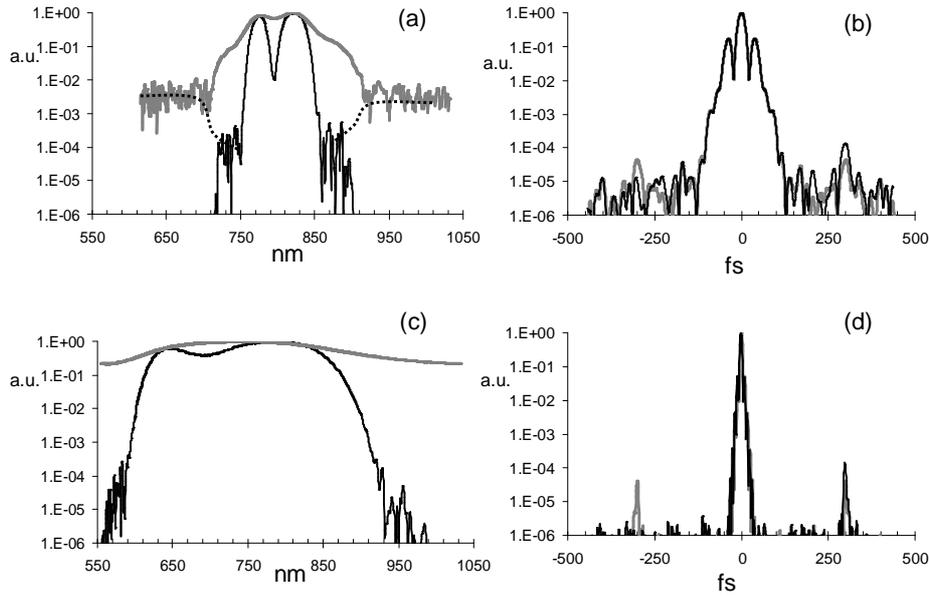

Fig.21: (a) XPW and input pulse spectral intensities for spectrum 2 with flat phase and a pulse replica of $10^{-4}$ at +300fs, (b) corresponding temporal intensities of the Fourier limit and SRSI measure of the input pulse, same quantities for spectrum 3 with flat phase and a pulse replica of $10^{-4}$ (c and d).

The dynamic difference from about 50dB for spectrum 2 to about 60dB for spectrum 3 corresponds to the filling factor difference.

It is very interesting to underline the noise diminution effect on the spectral intensity of the pulse directly due to the square roots difference and the larger spectral bandwidth of the XPW (dot line on fig.21.a).

The SNR factor of the input pulse spectrum is in fact better than the spectrometer one!

*9.3.2 Best SRSI device dynamic*

The spectrometer considered in this part is somehow the best that can be used. It has imaging properties without significant aberrations. The bandwidth can be adapted to fit the input pulse one. The dynamic of its cooled high dynamic CCD detector is 20000:1. The digitization is 16bits. As the CCD detector is an array of 2048 pixels per line and 512 lines. One can consider that without any deterioration, 100 lines can be averaged on a single shot. The optical setup uses best quality polarizer and optics to put the extinction ratio up to $10^6$.

The rule of thumb temporal contrast estimations are $C_t \approx 10^7 \cdot \sqrt{100} = 10^8 = 80 dB$ for spectrum 2 and $C_t \approx 5 \, 10^8 = 87 dB$ for spectrum 3.

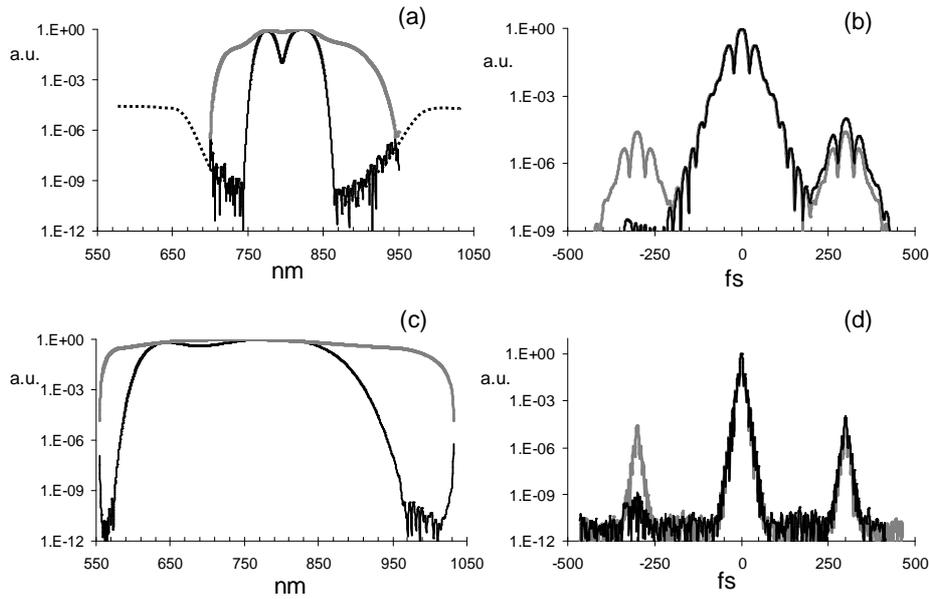

Fig.22 : (a) XPW and input pulse spectral intensities for spectrum 2 with flat phase and a pulse replica of $10^{-4}$ at +300fs, (b) corresponding temporal intensities of the Fourier limit and SRSI measure of the input pulse, same quantities for spectrum 3 with flat phase and a pulse replica of $10^{-4}$ (c and d).

The computer simulations present an even better dynamic by one order of magnitude. This is due to the self-heterodyne improvement on the spectrum intensity measurement of the pulse that has not been taken into account in the rule of thumb estimation.

With the 100 lines averaging, the highest simulated dynamic is more than $10^{10}$ in single shot mode.

## 10  Conclusions

In conclusion, the SRSI method is reviewed and simulated. The algorithm convergence and validity range are estimated from input pulse spectral intensity input. Measurement validity criteria from in situ data is explained and detailed. This ability of the measurement to be checked in situ make it very robust and easy to use.

The limitations of its experimental implementations can be overcome by adequate initial calibrations and data processing. This method can measure either picosecond pulse or sub-10fs pulse. It can be used from use to mid-IR as long as spectrometers are available. Its temporal dynamic is excellent and can fill the gap of high contrast single shot dynamic measurement in the few picosecond range.

## 11  Acknowledgements

The author thanks Daniel Kaplan, Antoine Moulet, Olivier Gobert, Vincent Crozatier, Mathias Herzog, Nicolas Forget, Stephanie Grabielle, Mathias Herzog, Sebastien Coudreau, Raman Maksimenka, Antoine Bourget, and for fruitful discussions and support. Fastlite acknowledges support from the Région Ile de France and the Conseil Général de l'Esssone.

## Appendix A: Notations and definitions

The real-valued electric field E(t) can be decomposed into monochromatic waves:

$$E(t) = \frac{1}{2\pi} \int_{-\infty}^{\infty} \tilde{E}(\omega) e^{i\omega t} d\omega \quad (1).$$

As E(t) real, $\tilde{E}(\omega)$ is hermitian. Thus the positive frequency part of the spectral components is sufficient for a full characterization of the pulse. So this part is defined as:

$$\tilde{E}^+(\omega) = \begin{cases} \tilde{E}(\omega) & \text{for } \omega \geq 0 \\ 0 & \text{for } \omega < 0 \end{cases} \quad (2).$$

The complex-valued temporal function $E^+(t)$ contains only the positive frequency segment of the spectrum and is the Fourier inverse transform of $\tilde{E}^+(\omega)$ and inversely:

$$E^+(t) = \frac{1}{2\pi}\int_{-\infty}^{\infty} \tilde{E}^+(\omega) e^{i\omega t} d\omega = FT^{-1}\left[\tilde{E}^+(\omega)\right]$$

$$\tilde{E}^+(\omega) = \int_{-\infty}^{\infty} E^+(t) e^{-i\omega t} d\omega = FT\left[E^+(t)\right] \quad (3).$$

The complex positive-frequency part $\tilde{E}^+(\omega)$ and $E^+(t)$ can be decomposed into amplitude and phase:

$$\tilde{E}^+(\omega) = \left|\tilde{E}^+(\omega)\right| e^{-i\phi(\omega)} = \sqrt{\frac{\pi I(\omega)}{\varepsilon_0 c n}} e^{-i\phi(\omega)}$$

$$E^+(t) = \left|E^+(t)\right| e^{i\Phi(t)} = \sqrt{\frac{I(t)}{2\varepsilon_0 c n}} e^{i\Phi(t)} \quad (4),$$

where I(t) is the temporal intensity and $I(\omega)$ is the spectral intensity proportional to the power spectrum density (PSD) – the familiar quantity measured with a spectrometer.

The spectral phase is often expanded into a Taylor series around a center frequency $\omega_0$:

$$\phi(\omega) = \sum_{j=0}^{\infty} \frac{\phi^{(j)}(\omega_0)}{j!}(\omega-\omega_0)^j \text{ with } \phi^{(j)}(\omega_0) = \left.\frac{\partial^j \phi(\omega)}{\partial \omega^j}\right|_{\omega_0} \quad (5).$$

The spectral phase coefficient of "null" order describes in the time domain the absolute phase. The first order term leads to a temporal translation of the envelope of the laser pulse in the time domain. A positive $\phi^{(1)}(\omega_0)$ corresponds to a shift toward later times. These two terms do not change the temporal structure of the pulse. Indeed only the coefficients of higher order are responsible for such changes. The second order term is commonly named chirp.

There is a variety of analytical pulse shapes where expressions in both domains remain analytical. Among them, Gaussian pulses are specific because its shapes remains Gaussian in both domain even with a pure chirp, and that any power of a Gaussian is still a Gaussian.

Before calculating the temporal filtering effect, Gaussian laser pulse characteristics need to be defined.

Let's consider perfectly compressed Gaussian pulse :

$$E^+(t) = \frac{E_0}{2} e^{-2\ln 2 \frac{t^2}{\Delta t_0^2}} e^{i\omega_0 t} \propto e^{-2\ln 2 \frac{t^2}{\Delta t_0^2}} = e^{-\frac{t^2}{2\Delta \tau_0^2}} \quad (6),$$

where $\Delta t_0$ is the full-width-at-half-maximum of the intensity, $\Delta \tau_0$ the statistical duration of the pulse defined as twice the standard deviation: $\Delta \tau_0 = \sigma_t = \sqrt{\langle t^2 \rangle - \langle t \rangle^2} = \sqrt{\int_{-\infty}^{\infty}(t-t_0)^2 \left|\tilde{E}^+(t-t_0)\right|^2 dt \bigg/ \int_{-\infty}^{\infty} \left|\tilde{E}^+(t-t_0)\right|^2 dt}$ (7), with

$t_0$ pulse center $t_0 = \langle t \rangle = \int_{-\infty}^{\infty} t\left|E^+(t)\right|^2 dt \bigg/ \int_{-\infty}^{\infty} \left|E^+(t)\right|^2 dt$ (8).

In the spectral domain, the same quantities can be defined:

$$\Delta\omega_0 = \sigma_\omega = \sqrt{\langle\omega^2\rangle - \langle\omega\rangle^2} = \sqrt{\int_{-\infty}^{\infty}(\omega-\omega_0)^2\left|\tilde{E}^+(\omega-\omega_0)\right|^2\frac{d\omega}{2\pi} \Big/ \int_{-\infty}^{\infty}\left|\tilde{E}^+(\omega-\omega_0)\right|^2\frac{d\omega}{2\pi}} \quad (9) \quad \text{and}$$

$$\omega_0 = \langle\omega\rangle = \int_{-\infty}^{\infty}\omega\left|\tilde{E}^+(\omega)\right|^2\frac{d\omega}{2\pi} \Big/ \int_{-\infty}^{\infty}\left|\tilde{E}^+(\omega)\right|^2\frac{d\omega}{2\pi} \quad (10).$$

The statistical widths are named rms duration and rms bandwidth in this article.
The spectrum of the Gaussian pulse is then:

$$\tilde{E}^+(\omega) = \frac{E_0 \Delta t_0}{2}\sqrt{\frac{\pi}{2\ln 2}}e^{-\frac{\Delta t_0^2}{8\ln 2}(\omega-\omega_0)^2} \propto e^{-\frac{\Delta t_0^2}{8\ln 2}(\omega-\omega_0)^2} = e^{-2\ln 2\frac{(\omega-\omega_0)^2}{\Delta\omega_{1/2}^2}} = e^{-\frac{(\omega-\omega_0)^2}{2\Delta\omega_0^2}} \quad (11).$$

By the more, the statistical duration can be expressed as a quadratic sum of two terms:

$$\Delta\tau^2 = \langle t^2\rangle - \langle t\rangle^2 = \int_{-\infty}^{+\infty}(t-t_0)^2 I(t-t_0)dt = \int_{-\infty}^{+\infty}TF\left[(t-t_0)E^+(t-t_0)\right]TF\left[(t-t_0)E^+(t-t_0)^*\right]dt$$

$$= \int_{-\infty}^{+\infty}\frac{\partial \tilde{E}^+}{\partial\omega}\frac{\partial \tilde{E}^{+*}}{\partial\omega}d\omega = \int_{-\infty}^{+\infty}\frac{\partial|\tilde{E}^+|}{\partial\omega}\frac{\partial|\tilde{E}^+|}{\partial\omega}d\omega + \int_{-\infty}^{+\infty}|\tilde{E}^+|^2\left(\frac{\partial\phi}{\partial\omega}\right)^2(\omega-\omega_0)^2 d\omega = \Delta\tau_0^2 + \Delta\tau_g^2$$

(12).

The first term $\Delta\tau_0$ is the minimal duration of the pulse said Fourier transform pulse or Fourier transform limited. The second term is linked to the spectral phase and the group delay by:

$$\Delta\tau_g = \sqrt{\left\langle(\tau_g - \langle\tau_g\rangle)^2\right\rangle} = \sqrt{\left\langle\left(\frac{\partial\phi}{\partial\omega} - \left\langle\frac{\partial\phi}{\partial\omega}\right\rangle\right)^2\right\rangle} \quad (13) \text{ with } \tau_g = \frac{\partial\phi}{\partial\omega} \text{ is the group delay.}$$

This relation is rigorous for any pulse shape but only for the statistical duration. For Gaussian pulses, the statistical and FWHM durations are proportional:

$$\Delta t = \frac{\Delta\tau}{\sqrt{8\ln 2}}, \Delta\omega_{1/2} = \frac{\Delta\omega}{\sqrt{8\ln 2}}.$$

So the same relation is obtained for FWHM duration.

In the same manner, the time-bandwidth product rigorously defined only statistically by $\Delta\omega\Delta\tau = \sigma_\omega\sigma_t \geq \alpha$ with $\alpha$ a constant dependant upon the shape of the pulse. $\alpha = 1/2$ for Gaussian pulses (its minimum value).

If there is no distortions due to high order spectral phase, $\Delta\tau_g = 0$ and $\Delta\omega_0\Delta\tau_0 = \alpha$.

By opposition, a purely chirped Gaussian pulse is stretched in time as:

$$\tilde{E}^+(\omega) \propto e^{-\frac{(\omega-\omega_0)^2}{2\Delta\omega_0^2}}e^{-i\frac{1}{2}\phi^{(2)}(\omega_0)(\omega-\omega_0)^2}$$

(14),

$$E^+(t) \propto e^{-\frac{t^2}{2\Delta\tau^2}}e^{i(at^2-\Phi_0)} = e^{-\frac{t^2}{2\Delta\tau_0^2\left(1+\left(\alpha\frac{\phi^{(2)}(\omega_0)}{\Delta\tau_0^2}\right)^2\right)}} e^{i\left(2\Delta\tau_0^4\left(1+\left(\alpha\frac{\phi^{(2)}(\omega_0)}{\Delta\tau_0^2}\right)^2\right)t^2-\Phi_0\right)}$$

where $\Delta\tau = \sqrt{\Delta\tau_0^2 + \Delta\tau_g^2} = \Delta\tau_0\sqrt{1+\left(\alpha\frac{\phi^{(2)}}{\Delta\tau_0^2}\right)^2}$ is the statistical duration of the chirped pulse.

This formulation is not restricted to Gaussian shape pulses.

## Appendix B: Noise in SRSI

The SRSI method have intrinsically a high temporal dynamic compared to the spectrometer dynamic. This appendix details the noise estimation and self-heterodyne nature of this measurement.

The interferogram with noise component writes [28]:

$\tilde{S}(\omega) = \tilde{S}_0(\omega) + \tilde{f}(\omega) e^{i\omega\tau} + \tilde{f}^*(\omega) e^{-i\omega\tau} + \tilde{\mathcal{N}}(\omega)$ (1), where $\tilde{\mathcal{N}}(\omega)$ is the offset white noise of the spectrometer.

In the time domain, the two signal are filtered with an identical gate. It results:

$$S_0(t) = (S(t) + \mathcal{N}(t)) \cdot G(t)$$
$$f(t) = (S(t) + \mathcal{N}(t)) \cdot G(t-\tau)$$ (2),

$$\tilde{S}_0(\omega) = \tilde{S}_0(\omega) \otimes \tilde{G}(\omega) + \tilde{\mathcal{N}}(\omega) \otimes \tilde{G}(\omega)$$
$$\tilde{f}(\omega) = \tilde{f}(\omega) \otimes \tilde{G}(\omega) e^{-i\omega\tau} + \tilde{\mathcal{N}}(\omega) \otimes \tilde{G}(\omega) e^{-i\omega\tau}$$ (3).

To recover without ambiguities the spectra, and we suppose that: $\tilde{S}_0(\omega) > |\tilde{f}(\omega)|$.

By extension of this hypothesis, we suppose that $\tilde{S}_0(\omega) >> |\tilde{f}(\omega)|$.

To simplify expressions (3), we suppose that $\tilde{G}(\omega) \approx \delta(\omega)$:

$$\tilde{S}_0(\omega) = \tilde{S}_\varnothing(\omega) + \tilde{\mathcal{N}}_S(\omega)$$
$$|\tilde{f}(\omega)| = |\tilde{f}(\omega)|_\varnothing + \tilde{\mathcal{N}}_f(\omega)$$ (4), where the subscript $\varnothing$ means noiseless.

Thus through the algebraic equations to recover the XPW and input spectral amplitude it comes out by using a limited development in $\tilde{\mathcal{N}}_S(\omega) \approx \tilde{\mathcal{N}}_f(\omega) \approx \tilde{\mathcal{N}}$:

$$\left|\tilde{E}^+_{XPW}(\omega)\right|^2 = \frac{1}{2}\left(\sqrt{\tilde{S}_0 + 2|\tilde{f}|} + \sqrt{\tilde{S}_0 - 2|\tilde{f}|}\right)^2 \approx \left|\tilde{E}^+_{XPW}\right|^2_\varnothing + \tilde{\mathcal{N}}\left(\frac{\left|\tilde{E}^+_{XPW}\right|_\varnothing}{B}\sqrt{\tilde{S}_\varnothing}\right) + o(\tilde{\mathcal{N}}^2)$$ (5) and

$$\left|\tilde{E}^+(\omega)\right|^2 = \frac{1}{4}\left(\sqrt{\tilde{S}_0 + 2|\tilde{f}|} - \sqrt{\tilde{S}_0 - 2|\tilde{f}|}\right)^2$$
$$\approx \left|\tilde{E}^+\right|^2_\varnothing + \tilde{\mathcal{N}}\left(\frac{\left|\tilde{E}^+\right|_\varnothing}{B}\sqrt{\tilde{S}_\varnothing}\right) + \tilde{\mathcal{N}}_f^2\left(\frac{B^4 - 4\tilde{S}_\varnothing^2}{2B^3}\right) - 2\tilde{\mathcal{N}}_f^2\left(\frac{B^4 - 4|\tilde{f}|^2_\varnothing}{B^3}\right) + \tilde{\mathcal{N}}_S\tilde{\mathcal{N}}_f\left(\frac{4\tilde{S}_\varnothing|\tilde{f}|_\varnothing}{B^3}\right) + o(\tilde{\mathcal{N}}_S^3, \tilde{\mathcal{N}}_f^3)$$ (6),

where the subscript $\varnothing$ means noiseless, and $B = \sqrt{\tilde{S}_0^2 - 4|f|^2_\varnothing}$.

As the XPW spectrum is larger than the input spectrum, it is stronger than the input signal on the sides:
$$\left|\tilde{E}^+_{XPW}\right| >> \left|\tilde{E}^+\right| \Leftrightarrow \tilde{S}_\varnothing >> 2|\tilde{f}|_\varnothing \Rightarrow B \approx \tilde{S}_\varnothing \approx \left|\tilde{E}^+_{XPW}\right|^2_\varnothing$$

Then the two expressions can be simplified as:

$$\left|\tilde{E}^+_{XPW}\right|^2 \approx \left|\tilde{E}^+_{XPW}\right|^2_\varnothing + \tilde{\mathcal{N}} + o(\tilde{\mathcal{N}}^2)$$ (7)

$$\left|\tilde{E}^+(\omega)\right|^2 \approx \left|\tilde{E}^+\right|^2_\varnothing + \tilde{\mathcal{N}}\left(\frac{\left|\tilde{E}^+\right|_\varnothing}{\left|\tilde{E}^+_{XPW}\right|_\varnothing}\right) + \tilde{\mathcal{N}}^2\left(2\left|\tilde{E}^+_{XPW}\right|^2_\varnothing\right) + o(\tilde{\mathcal{N}}^3)$$ (8).

The signal-to-noise ratio of the two spectra are then given by:

$$SNR\left(\left|\tilde{E}^+_{XPW}(\omega)\right|^2\right) \approx \frac{\left|\tilde{E}^+_{XPW}(\omega)\right|^2_\varnothing}{\tilde{\mathcal{N}}}$$ (9)

$$SNR\left(\left|\tilde{E}^+(\omega)\right|^2\right) \approx \frac{\left|\tilde{E}^+_{XPW}\right|_\varnothing \left|\tilde{E}^+\right|_\varnothing}{\tilde{\mathcal{N}}\left(1 + 2\frac{\left|\tilde{E}^+_{XPW}\right|_\varnothing}{\left|\tilde{E}^+\right|_\varnothing}\tilde{\mathcal{N}}\right)} \quad (10).$$

The SNR of the XPW pulse is unchanged compared to a direct measurement by the spectrometer.

But for the input pulse, the SNR is amplified by the the ratio of the two spectral amplitude : $\frac{\left|\tilde{E}^+_{XPW}\right|_\varnothing}{\left|\tilde{E}^+\right|_\varnothing}$ to the first order in noise contribution. This amplification leads to few orders improvement on the spectrum dynamics compared to a direct measurement by the same spectrometer as shown on figures 21 and 22.

The self-heterodyne detection for the input spectrum $\left|\tilde{E}^+(\omega)\right|^2$ approximately decreases the noise contribution by the dynamic of the XPW spectrum $\frac{\left|\tilde{E}^+_{XPW}\right|_\varnothing}{\left|\tilde{E}^+\right|_\varnothing}$ on the side of the spectrum.

**Appendix C: Offset white noise without relevant photon or shot noise**

The relevant noise in SRSI method is an offset white noise on the interferogram. This noise is the result of the dark current noise from the CCD pixels, and the read-out electronics Johnson type noise. It is characterized by the dynamic of the CCD detector device:

$$Dynamic = SNR = \frac{Max(Signal)}{Noise} .$$

The shot noise or photon noise is considered as not relevant in here. This assumption deserves some explanations.
As demonstrated and illustrated in [22], "the photon noise is related to the statistical fluctuation of the photons collected by the photodetector. The photon counting statistics is known to be Poissonian: if the detector surface receives N photons in average during an integration time t, the standard deviation of the number of photons received is $\sqrt{N}$ . The photoelectrons created in the detector obey the same Poissonian statistics and this explains the shot-noise on the photocurrent". Ordinary light sources are at the shot-noise limit. Some light sources are well above this limit. Some other are below this limit as proven by sub shot noise measurements. The photodetector is not responsible for the shot-noise. Shot-noise is related to the quantum "nature" of the light and the shot-noise limit is due to the Poissonian statistics of the photon collected.
In the SRSI method, the noise to consider is the noise on the CCD pixels of the spectrometer. The mode-locked nature of the ultrafast laser pulses completely breaks down the Poissonian statistics of the photon. The photons measured in SRSI are not randomly generated in time but respect a temporal constraint due to the mode-lock. This constraint makes the photon source sub shot noise for the spectral intensity.
As an example, if photon noise applies to this kind of pulses, then the temporal intensity of the Fourier limit should be limited in contrast to about $10^6$ as shown in Fig.23 below. Oscillators and last generation of CPA lasers have contrast in the range of $10^8$ to $10^{10}$ proving the this noise is linked to the temporal profile of the pulse and not to an inherent characteristic of the light.

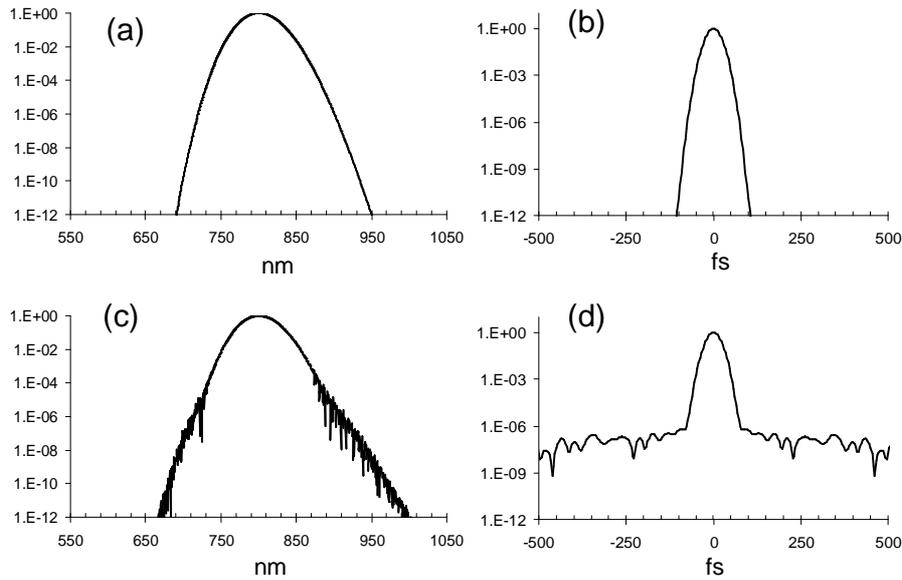

Fig.23: illustration of sub shot noise nature of a mode-lock pulse spectral intensity measurement: (a) spectral intensity without shot noise, (b) corresponding temporal intensity, (c) spectral intensity with shot noise, (d) corresponding temporal intensity.